\documentclass{article}

\usepackage{arxiv}

\usepackage[utf8]{inputenc} 
\usepackage[T1]{fontenc}    
\usepackage{hyperref}       
\usepackage{url}            
\usepackage{booktabs}       
\usepackage{amsfonts}       
\usepackage{nicefrac}       
\usepackage{microtype}      
\usepackage{lipsum}
\usepackage{graphicx}
\usepackage{subfigure}

\title{Nonparametric sign prediction of high-dimensional correlation matrix coefficients}

\author{
 Christian Bongiorno\thanks{} \\
  Laboratoire de Math\'ematiques et Informatique pour les Syst\`emes Complexes,\\
   CentraleSup\'elec, Universit\'e Paris Saclay,\\
    3 rue Joliot-Curie, 91192, Gif-sur-Yvette, France\\
  \texttt{christian.bongiorno@centralesupelec.fr} \\
   \And
 Damien Challet \\
  Laboratoire de Math\'ematiques et Informatique pour les Syst\`emes Complexes,\\
   CentraleSup\'elec, Universit\'e Paris Saclay,\\
    3 rue Joliot-Curie, 91192, Gif-sur-Yvette, France\\
}

\begin{document}
\maketitle


%
%

\begin{abstract}
We introduce a method to predict which correlation matrix coefficients are likely to change their signs in the future in the high-dimensional regime, i.e. when the number of features is larger than the number of samples per feature. The stability of correlation signs, two-by-two relationships, is found to depend on three-by-three relationships inspired by Heider social cohesion theory in this regime. We apply our method to US and Hong Kong equities historical data to illustrate how the structure of correlation matrices influences the stability of the sign of its coefficients.
\end{abstract}


\flushbottom
\maketitle
%
%
\thispagestyle{empty}


\section*{Introduction}

Correlation matrices may be pathologically noisy without a proper filtering method. For example building optimal mean-variance portfolios \cite{markowitz1952portfolio} requires so precise estimations of trends, covariances and correlations that portfolio optimization was equated to error maximization by Michaud\cite{michaud1989markowitz}. The main problem is that a precise estimation of a full unfiltered correlation matrix between $N$ features requires $T\gg N $ samples  per feature. Regrettably, the non-stationary nature of many real-life dynamical systems, including financial markets, imposes $T$ to be as small as reasonably possible but in any case proportional to $N$. The impossibility to approximate the $T\to\infty$ limit while keeping $N$ constant is known as the `curse of dimensionality' as correlation estimators remain noisy even in the $N$ and $T\to\infty$ limit at fixed ratio $q=T/N$. 

Ad-hoc filtering techniques include linear shrinkage\cite{ledoit2004honey}, block-diagonal ansatz for the correlation matrix\cite{marsili2002dissecting} and random matrix theory-based eigenvalue clipping\cite{laloux1999noise,plerou2002random}. The latter works reasonably well for $T>N$. More recently, the Rotational Invariant Estimator (RIE), which makes use of eigenvectors as well, was shown to be optimal in the large $N$ and $T$ limit at constant ratio $q=T/N>1$\cite{bun2016rotational}. RMT and RIE assume stationary Gaussian returns and $N<T$ (see Bun et al~\cite{bun2017cleaning} for a recent review of the field). 
Here, we focus on non-stationary correlation structures of possibly non-Gaussian returns when $N>T$ and aim to predict the sign of asset correlations.  Statistics calls the $N>T$ case high-dimensional and we will follow this terminology.

Here, we are considering a complex network representation of an asset correlation matrix. Historically, the first use of complex networks in finances is the Minimum Spanning Tree (MST)~\cite{mantegna1999hierarchical}. In this work, the author showed that a tree network could adequately describe the economic sector taxonomy of a portfolio of $N$ assets,  leading to a substantial complexity reduction, i.e., from $N \times N$ correlation coefficients to $N-1$ links. Later on, in Aste et al\cite{aste2005complex} and Tuminello et al\cite{tumminello2005tool}, the authors proposed a relaxation of the topological constraint of the MST, leading to the Maximal Planar Graph, i.e., a filtered graph embedded in a bi-dimensional space. In this work, we use the most straightforward parametric procedure to obtain a network from a correlation matrix, which is the asset graph\cite{onnela2004clustering}.  The asset graph prescribes to retain only links related to correlations exceeding a threshold value. 

 Whereas correlations involve two time series (they are dyadic), we find that triadic measures better quantify the global and local stability of the dependence and thus better predict the stability of the sign of correlations when $T<N$. 
Our approach is related to Heider balance theory\cite{heider1946attitudes,cartwright1956structural} which aims at explaining the attitude changes of interacting individuals. In the modeling framework of this theory, only two possible interactions between two individuals are possible: the latter can be \emph{friends} or \emph{enemies}. The general observation in social science  that `the enemy of my friend is my enemy'\cite{rapoport1963mathematical} becomes particularly relevant when extended to triadic relationships: for example, triads where $a$ is a friend of $b$ and $c$ but $c$ is an enemy of $b$ tend to be unstable. As a consequence, one interaction type is likely to change and lead to a stable triad: $a$ could become an enemy of $b$, or $c$ could become a friend of $b$. In a similar way, a triad composed of three individuals that are enemies of each other is considered unstable as two individuals could join their forces against the third one. In summary, this theory identifies four possible triads, two stable ones and two unstable ones, and adds the intuition is that unstable triads tend to evolve into stable ones.

More recently, many authors in the field of network science proposed to extend the mechanism of triad balance to describe the evolution of a signed complex network\cite{altafini2012dynamics,antal2006social,leskovec2010signed,bianconi2014triadic}. In particular, Hedayatifar \cite{hedayatifar2017pseudo} measured the global social balance with a Hamiltonian whose minimal energy level coincides with the maximal stability and studied the possible paths that drive the system towards minimal energy levels, i.e. to the maximally stable triad states. In a financial context, various properties of network structures have been used to characterize the state of the market \cite{onnela2005intensity,aste2010correlation} or of interbank lending networks\cite{haldane2011systemic,bardoscia2017pathways}.

Here, we link the stability of the dyadic relationships as encoded by correlation matrices and statistically validated networks to triadic relationships.

\section*{Results}
\subsection*{Data Description and Processing}\label{sec:data}
In this work we consider the daily close-to-close returns of equities from US and Hong Kong stock markets, adjusted for dividends, splits and other corporate events. We focus on large capitalization equities for which we have information on their official industrial sector. More precisely:
\begin{enumerate}
\item  US equities: large-capitalization stocks, from 1992-02-03 to 2018-06-29.  The number of stocks with data vary over time: it ranges from 399 in 1992-02-06 to 723 in 2018-06-29, and is roughly constant from 2008.
\item Hong Kong equities: 1277 stocks with the largest capitalization as of 2019-05-01 listed on Hong Kong stock exchange. Our dataset covers the 2002-01-04 to 2017-06-23 period. The number of stocks is similar to the US database: the minimum number of stocks was 320 stocks on 2002-01-01 and the maximum was 1277 on 2017-06-23.
\end{enumerate}

Let $p_{i,t}$ be the matrix of the adjusted close prices, we denote the log-return matrix $r$ whose elements are  $r_{i,t} = \log( p_{i,t}) -\log(p_{i,t-1})$.

First, we define partial returns $\tilde{r}_{i,t}=r_{i,t}-m_t$ where $m_t$ is the median of all price returns at time $t$ (a nonparametric definition of the market mode\cite{borghesi2007emergence}) and their binarized values $b_{i,t}=\textrm{sign}(\tilde{r}_{i,t})$. 

Then, for a given time window $[t-T+1,t]$, we only keep the assets without any missing value and evaluate the correlation matrix $\Phi_t$ of the binarized returns $b$\cite{ozer1985correlation}. Given the definition of $b$, $\Phi_t$ is nonparametric. We also will use the notation $C_t$ to denote the correlation matrix of the raw returns $r_{i,t}$ in the time window $[t-T+1,t]$.

Binarized returns certainly require less bits of storage. Whether they contain less information in practice depends on the situation and on the issue under investigation. For example, the cluster composition of US equities determined by Louvain clustering adapted to usual correlation matrices\cite{blondel2008fast} is essentially the same if for $\Phi$ and $C$\cite{almog2014binary,almog2015mesoscopic}. Here, we use binarized returns for two reasons: first to infer statistically validated networks, and second to build a robust nonparametric method.

For the sake of completeness (but not robustness), we repeated in the Supplementary Information all the analysis with Pearson correlation matrices computed of raw returns, and we achieved qualitatively similar results. In addition, we reported in the Supplementary Information an alternative way to remove the global trend from the Pearson correlation matrix.\\

\subsection*{Fast correlation structure dynamics}\label{sec:cluster}

We first illustrate how quickly the structure of correlation matrices change in financial markets, which explains why the prediction of correlation sign changes is important in this context. The idea is to infer statistically significant elements of $\Phi_t$, which then defines a time-dependent adjacency matrix $A_t$ whose evolution reflects some of the structural changes of the financial market in question. 

We restrict $\Phi_t$ to its significantly positive elements by controlling for multiple hypothesis. Because $\Phi_t$ is computed from binary variables, we can use the one-sided Fisher exact test, which equivalent to the hypergeometric test\cite{rivals2006enrichment}.    
 The coefficients which pass the test (at a false discovery rate set to $\alpha=0.1$) form the adjacency matrix $A_t$ whose coefficients are either $A_{ij,t}=1$ if $\phi_{ij,t}$ is selected or 0 otherwise. This is known as Statistically Validated Networks (SVNs)~which may be applied to more than two states\cite{tumminello2011statistically}. Section Methods gives for more details about the method. 
The main advantage of this approach is to obtain a filtered network of stocks even in the high-dimensional regime ($N>T$), without any assumption on the resulting clustering structure such as not-overlapping clusters and without resorting to bootstraps\cite{cai2016large,reigneron2019case}

It is well known that stocks belonging to the same sector are usually strongly correlated with each other: several methods can observe such emergent behavior, for example, the minimum spanning tree\cite{mantegna1999hierarchical} or principal component analysis coupled with random matrix theory\cite{laloux1999noise,plerou2002random}. Here, we expect that assets that belong in the same sector form clusters in the SVN $A_t$. Among the arsenal of techniques to screen a complex network, we identified in the assortativity coefficient\cite{newman2010networks}  as the most appropriate to measure the similarity with respect to the GICS sector composition. We shall drop the index $t$ when it leads to too heavy notations. The assortativity coefficient is defined as 
$$
G = \frac{\sum_{ij}\left(A_{ij}-k_i k_j/2m\right) \delta(c_i,c_j)}{2m-\sum_{ij}(k_i k_j/2m) \delta(c_i,c_j)}
$$
where $\delta(c_i,c_j)$ is $1$ if node $i$ and $j$ belong to the same sector and zero otherwise, $k_i$ is the degree of node $i$ i.e. $k_i = \sum_j A_{ij}$ and $m$ is the total number of links. 

By definition, assortativity $G \in [-1,1]$; its expectation equals $0$   if the links of the networks are distributed randomly with respect to the sector partition in the case of a configuration null-model. $G>0$ indicates a propensity of the nodes to establish links between nodes the same sector and reversely.

This section reports results for US equities; those for Hong Kong equities are to be found in SI. We explored the dynamics of $G$ by shifting a time window of $T=100$ days one day at a time from 2000-01-03 to 2018-06-29. For each time window $t$, we computed the SVN $A_t$  with $\alpha=0.1$ and the related assortativity coefficient $G_t$.
\begin{figure*}[tbh]
\centering
\subfigure[\label{fig:assort}]{\includegraphics[width=0.3\textwidth]{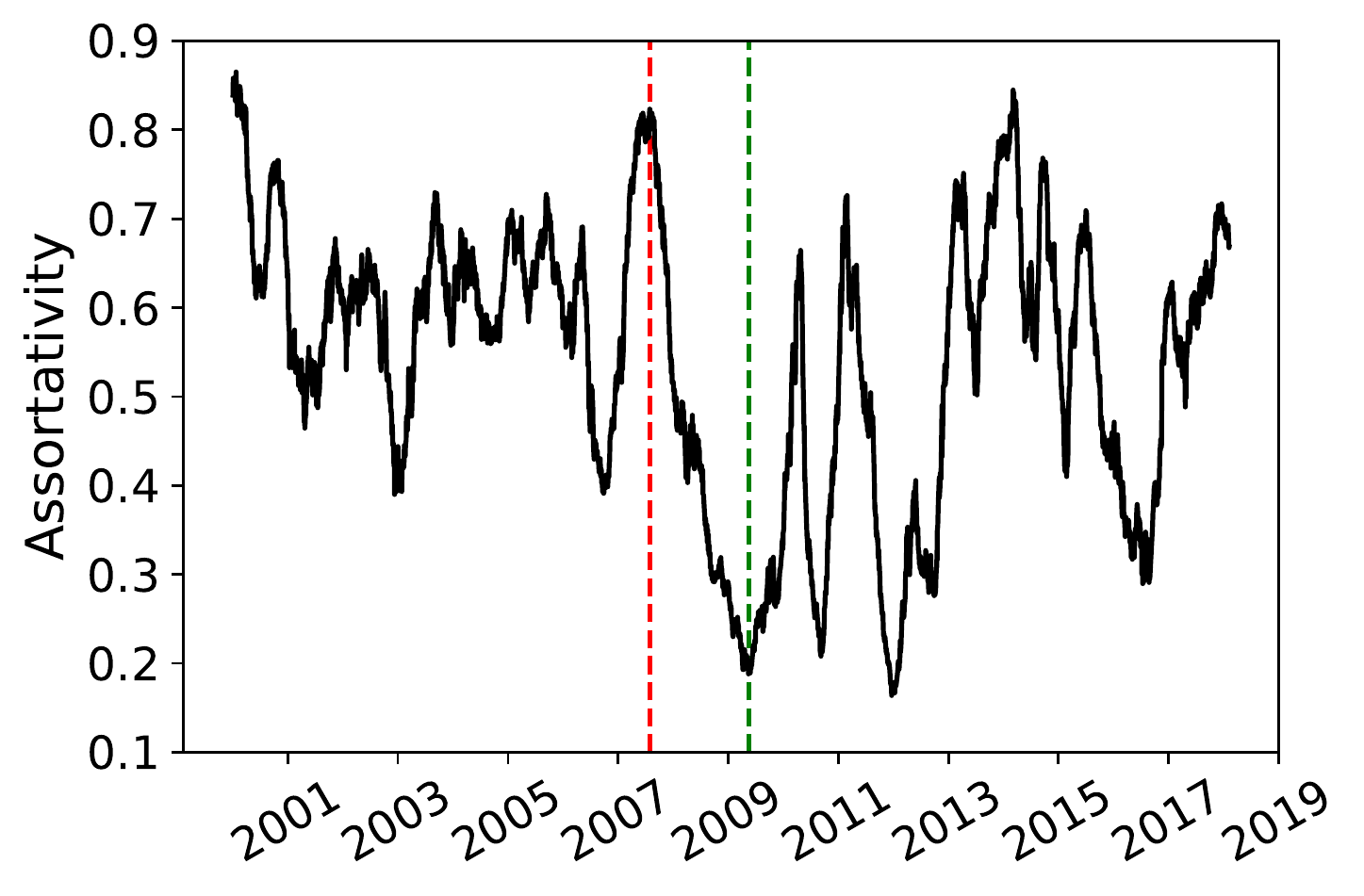} }
\subfigure[\label{fig:nlink}]{\includegraphics[width=0.3\textwidth]{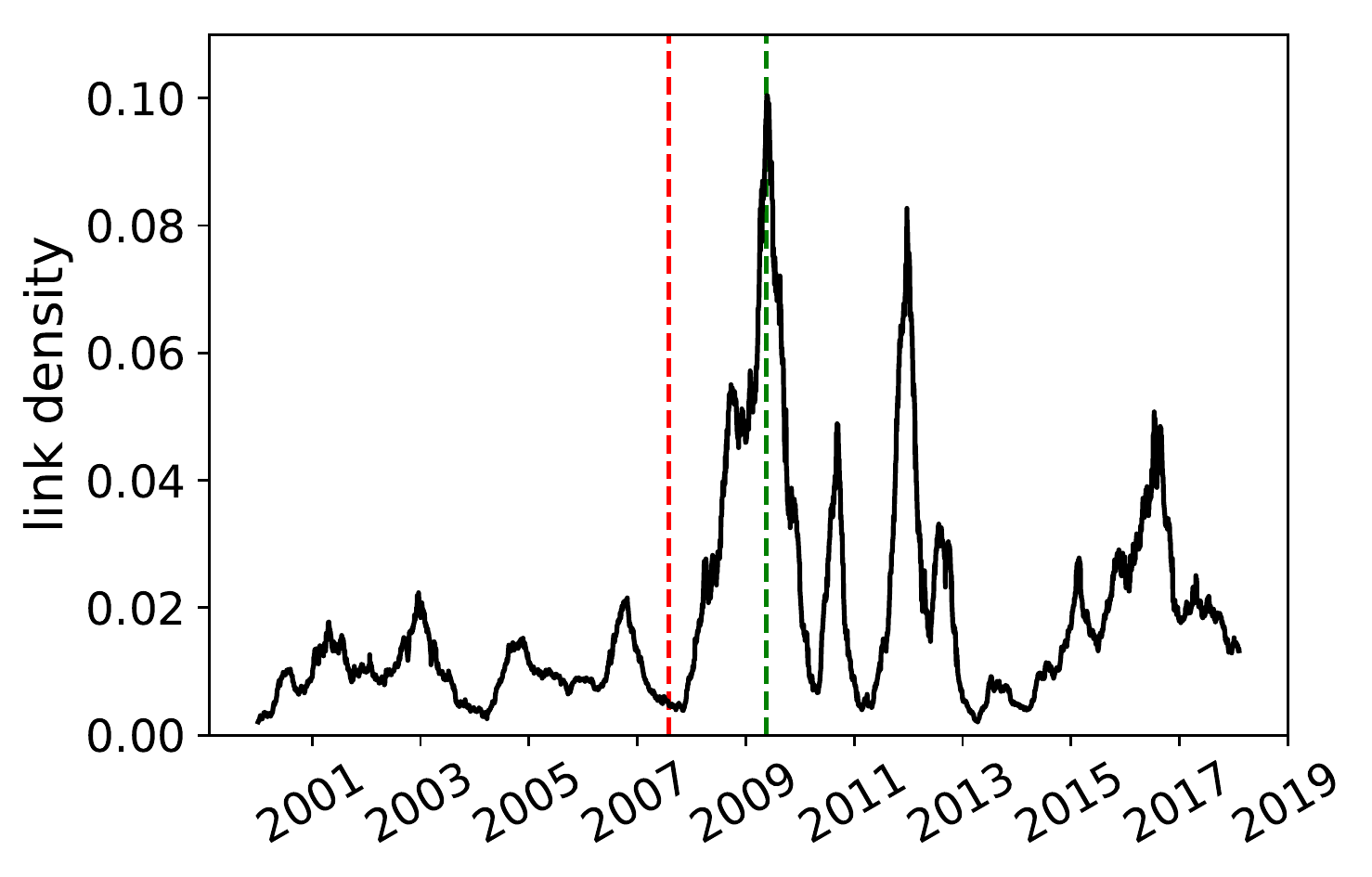} }

\subfigure[\label{fig:2007net}]{\includegraphics[width=0.3\textwidth]{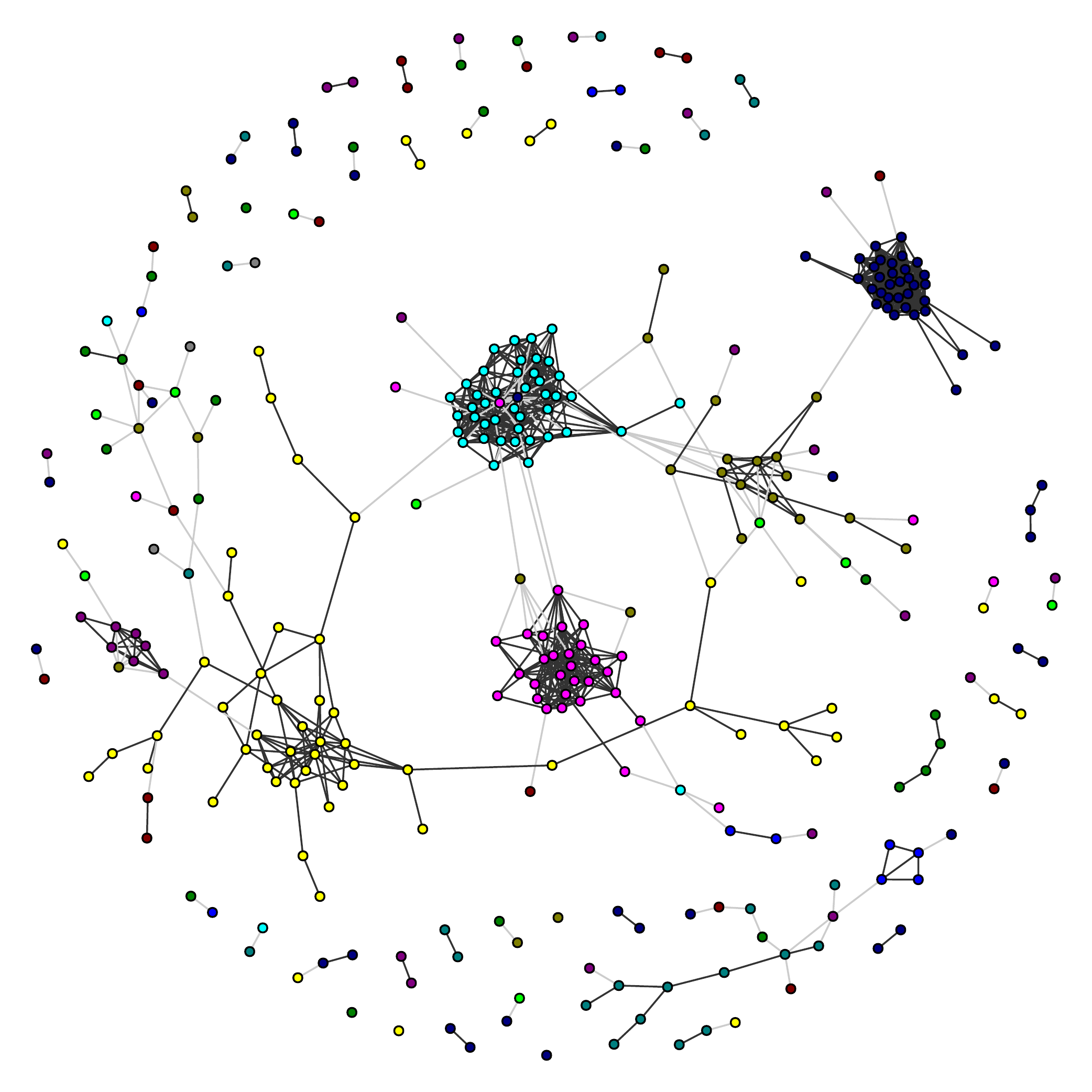} }
\subfigure[\label{fig:2009net}]{\includegraphics[width=0.3\textwidth]{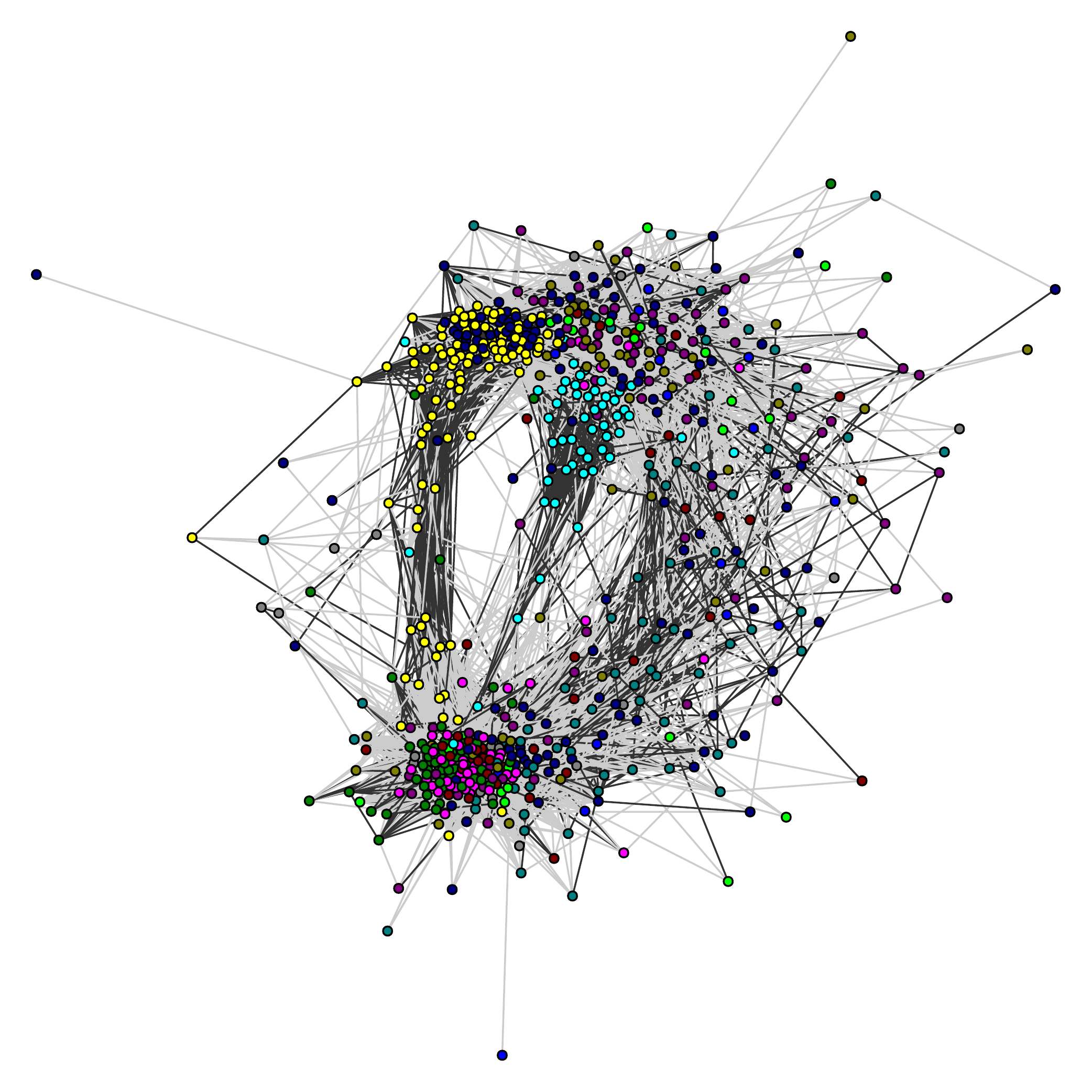} }
\caption{\small{$(a)$ Assortativity of the network of statistically validated correlation (SVN) with respect to the sector classification, $(b)$ links density of the correlation network, each point refers to the last day of the time window used to compute the SVN; $(c)$ correlation network  as of 2007-07-30 (dotted red line in panel (a)); $(d)$ correlation network as of 2009-05-19 (dotted green line in panel $(a)$); the color code of the nodes of both networks represents different different sectors, the links between different sectors are colored in gray; calibration windows of $T=100$ days.} }
\end{figure*}

The assortativity coefficient oscillates between $0.2$ and $0.85$ (Fig.~\ref{fig:assort}). One of the local minimum values is reached during the crisis of 2009 and is close to the random sector linking limit. It is worth noticing that although the link density of SVNs reaches local maxima in the proximity of the minima of the assortativity (Fig.~\ref{fig:nlink}), the dynamical evolution of the assortativity should be unbiased by definition with respect to the number of links as indeed the assortativity coefficient is an adjusted metric that considers the configuration model as a null model, i.e., the family of models that preserves the exact degree distribution of the nodes.

In order to support our interpretation of the assortativity coefficient, we show a graphical representation of the networks obtained in 2007-07-20 and 2019-05-19 in Figs~\ref{fig:2007net} and Fig.~\ref{fig:2009net} respectively.  The clusters observed in the network of Fig.~\ref{fig:2007net}, characterized by an assortativity of $0.8$, has a clear association with the macroscopic structure defined by the sectors. However, in the network of Fig.~\ref{fig:2009net}, such association disappears. Such network, characterized by an assortativity of $0.2$, seems to be composed by two large clusters, highly overlapping each other. 
Results for significantly negative correlations are reported in S.I.; in short, they lead to much sparser networks that are disassortative ($G<0$) and only non-null in times of crisis.

Thus SVNs of binarized partial returns are able to capture part of the fast  evolution of correlation structures in a way that overcomes the usual problems of correlation matrices in the high-dimensional regime $N>T$. That said, because SVNs are built by controling the false positive rate, they do not control the false negative rate, i.e,, the fraction of links that have been wrongly omitted (see Bongiorno et al\cite{bongiorno2017core} for a discussion on this point). The smaller the FDR of the SVN, the larger the risk of a larger false negative rate. Figure~\ref{fig:nlink} illustrates this point: only a few links are retained in the SVN for most of the time periods and a high number of nodes are isolated.

At any rate, it is clear that the structure of correlation has a non-trivial fast dynamics, which can only be captured by small calibration windows. In the following, we focus on a specific part of structural changes, i.e. the change of correlation signs.

\subsection*{Triads Dynamic}\label{sec:triads}

We first further simplify the correlation matrix $\Phi$ by taking its sign and by setting its diagonal to 0: we introduce $S=\textrm{sign}({\Phi})-\mathbb{I}$. In short, one assumes that a link is positive if $\Phi_{ij}\geq 0$ and a negative when $\Phi_{ij} < 0$, and  $S_{i,i}$=0. This time, an unknown fraction of false positives may be included in $S$. However, the global information emerging by considering the whole network structure will compensate for such errors.

\begin{figure*}
\centering
\includegraphics[width=1\textwidth]{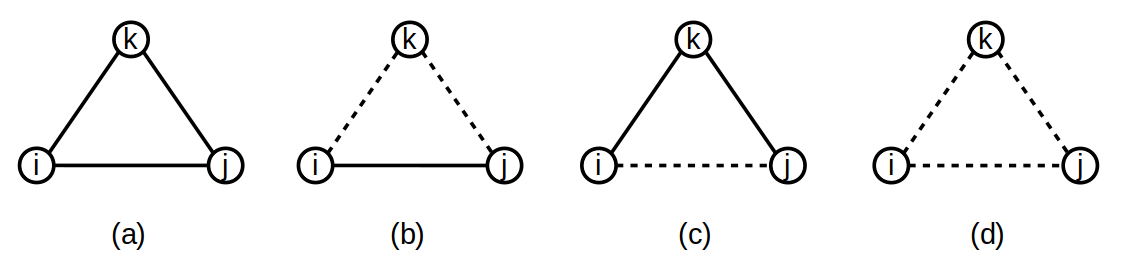}
\caption{\small{$(a)$ and $(b)$ stable triads, $(c)$ and $(d)$ unstable triads. The solid line indicates a positive relation, the dotted line indicates a negative relation.}}\label{fig:triads}
\end{figure*}

The matrix $S$ is nothing else than a signed adjacency matrix and makes it easy to define triads: there are four possible triads, two stable and two unstable (see Fig.~\ref{fig:triads}) ones. In the case of asset returns, the two stable configurations correspond to a triangle of positively correlated assets, and to two assets that are positively correlated but negatively correlated to a third one. The two unstable situations involve two positive links and three negative ones. 

Thus, triads with an odd number of negative links are stable and those with an even number of negative links are unstable. Hedayatifar et al\cite{hedayatifar2017pseudo} introduce a global metric $H$ to characterize the fraction of stable triangles in a system of $N$ nodes, defined as
\begin{equation}\label{eq:H1}
H = - \frac{1}{{N \choose 3}} \sum_{ijk} S_{ij} S_{ik} S_{jk}.
\end{equation}
A stable triangle adds $+1$ to the sum and an unstable one  $-1$. The metric is normalized by the maximum number of triangles. Finally the minus sign ensures that a system with only stable triangles has $H=-1$, and a system characterized by only unstable triangles has $H=+1$. Thus, as pointed out in Hedayatifar et al\cite{hedayatifar2017pseudo}, $H$ can be interpreted as the Hamiltonian of a physical system. Only two possible macroscopic states are possible in the lowest energy levels (the most stable ones): the `paradise', where all the nodes have positive links with other nodes, and the `bi-polar' with two groups with positive interactions within the same group and negative interactions among different groups. Other compositions such as a clustering structure with $K>2$ clusters can exist in a jammed state or be caused by an external force. 

Within this modelling framework, the clustering structure such as the sector composition may therefore to be unstable with respect to perturbations and may evolve towards more stable structures if it was not for stabilizing forces that we do not explicitly account for in the following. 

\begin{figure}
\centering
\subfigure[\label{fig:Hass}]{\includegraphics[width=7cm]{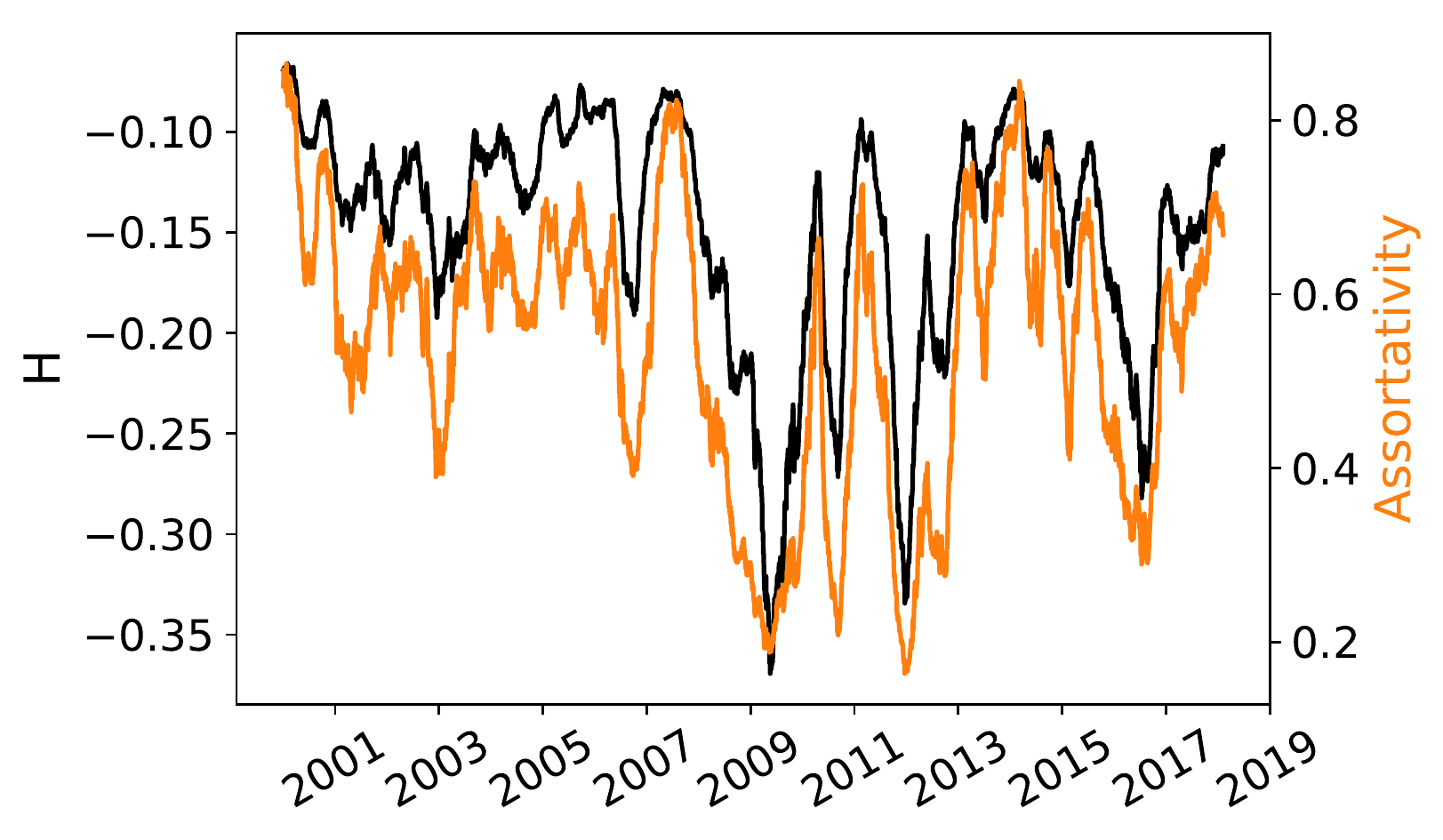}}
\subfigure[\label{fig:Hvol}]{\includegraphics[width=7cm]{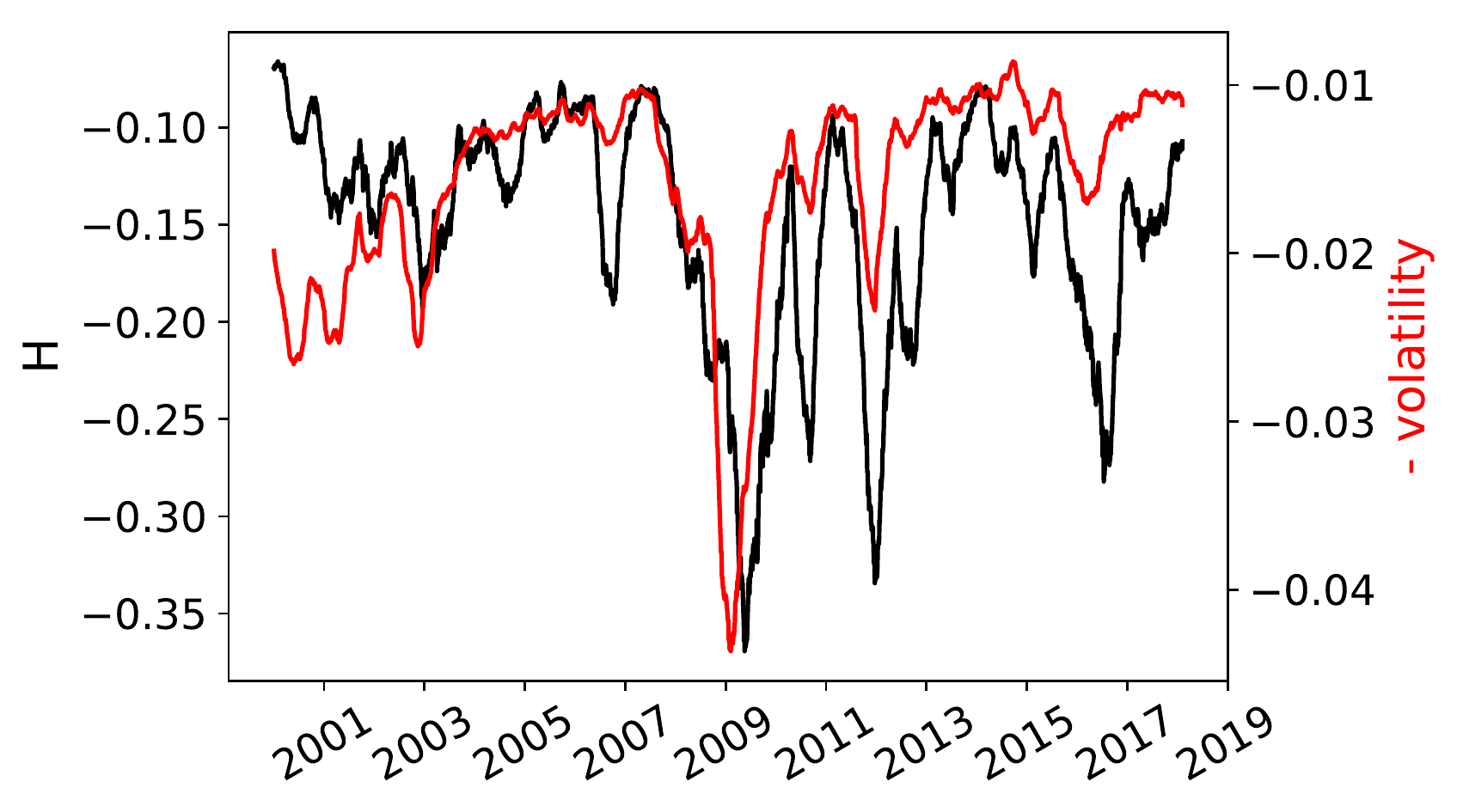}}

\caption{\small{$(a)$ Evolution of $H$ (black line), assortativity coefficient of the SVNs with respect to the sector composition (orange line); $(b)$ Evolution of $H$ (black line), minus volatility (red line) defined as the average absolute value of the returns in the considered time window.}}
\end{figure}

We observed the evolution of $H$ with a time window of $T=100$ days. As shown in Fig.~\ref{fig:Hass}, the dynamic of $H$
is strongly correlated with the sector assortativity of the SVNs. In fact, the local minima of the assortativity correspond to the local minima of $H$, and similarly for the local maxima. This observation confirms that such a dynamic of the composition of the clusters can be well detected by $H$. It is worth noticing that the evolution of $H$ is strongly anticorrelated with that of the volatility, see Fig.~\ref{fig:Hvol}. In particular, a large change of the volatility in many cases precedes a similar event for $H$, as in 2009. This suggests that the volatility could be related with the perturbation that moves the system away from the jammed state characterized by the sector structure.

\subsubsection*{Triads and Spectral Decomposition}
According to the Spectral Decomposition Theorem, a symmetric matrix can be written as a sum of its eigenvectors weighted by their respective eigenvalues
\begin{equation}
\Phi  = \sum_{i=1}^{N} \lambda_i \mathbf{v}_i'\,\mathbf{v}_i,\label{eq:rho_v_i}
\end{equation}
where $\lambda_i$ is the $i$-th eigenvalue, $\mathbf{v}_i$ its associated eigenvector, and $\mathbf{v}_i'$ the transpose of $\mathbf{v}_i$. In addition, since correlation matrices are positive-defined, $\lambda_i\ge 0$ for every $i$. When $q=T/N<1$, there are $N-T$ zero eigenvalues and the rank of the matrix is $T$, hence smaller than $N$. Finally, $\sum_i \lambda_i =N$. We shall adopt the convention that eigenvalues are sorted, i.e., $\lambda_1\ge\lambda_2\ge\cdots\ge\lambda_N$.

This decomposition yields interesting insights on triads: indeed, each component of the spectral decomposition $\mathbf{v}_i'\,\mathbf{v}_i$ is a matrix which only contains stable triangles. This means that the signs of its elements label the groups to which they belong. Indeed, only two possible scenarios can occur: 

\begin{enumerate}

\item {\bf paradisiac case}: $\mathbf{v}_i$ has only positive (or negative) components, in which case $\mathbf{v}_i'\,\mathbf{v}_i$ has only positive entries;
\item {\bf bi-polar case}: some components of $\mathbf{v}_i$ are positive and others are negative, in which case the matrix $\mathbf{v}_i'\,\mathbf{v}_i$ is composed of two groups.
\end{enumerate}

Therefore, if the largest eigenvalue is much larger than the other ones, i.e. $\lambda_1 \gg 1/N$, the  $\textbf{v}_1'\,\textbf{v}_1$ matrix dominates in Eq.\ (\ref{eq:rho_v_i}). Reversely, if $\lambda_1 \approx 1/N$, the contribution to the stability of each component  $\textbf{v}_i'\,\textbf{v}_i$ may cancel out each other, leading in most of the cases to lower global stability.

\begin{figure}
\centering
\subfigure[\label{fig:lmdH}]{\includegraphics[width=0.4\textwidth]{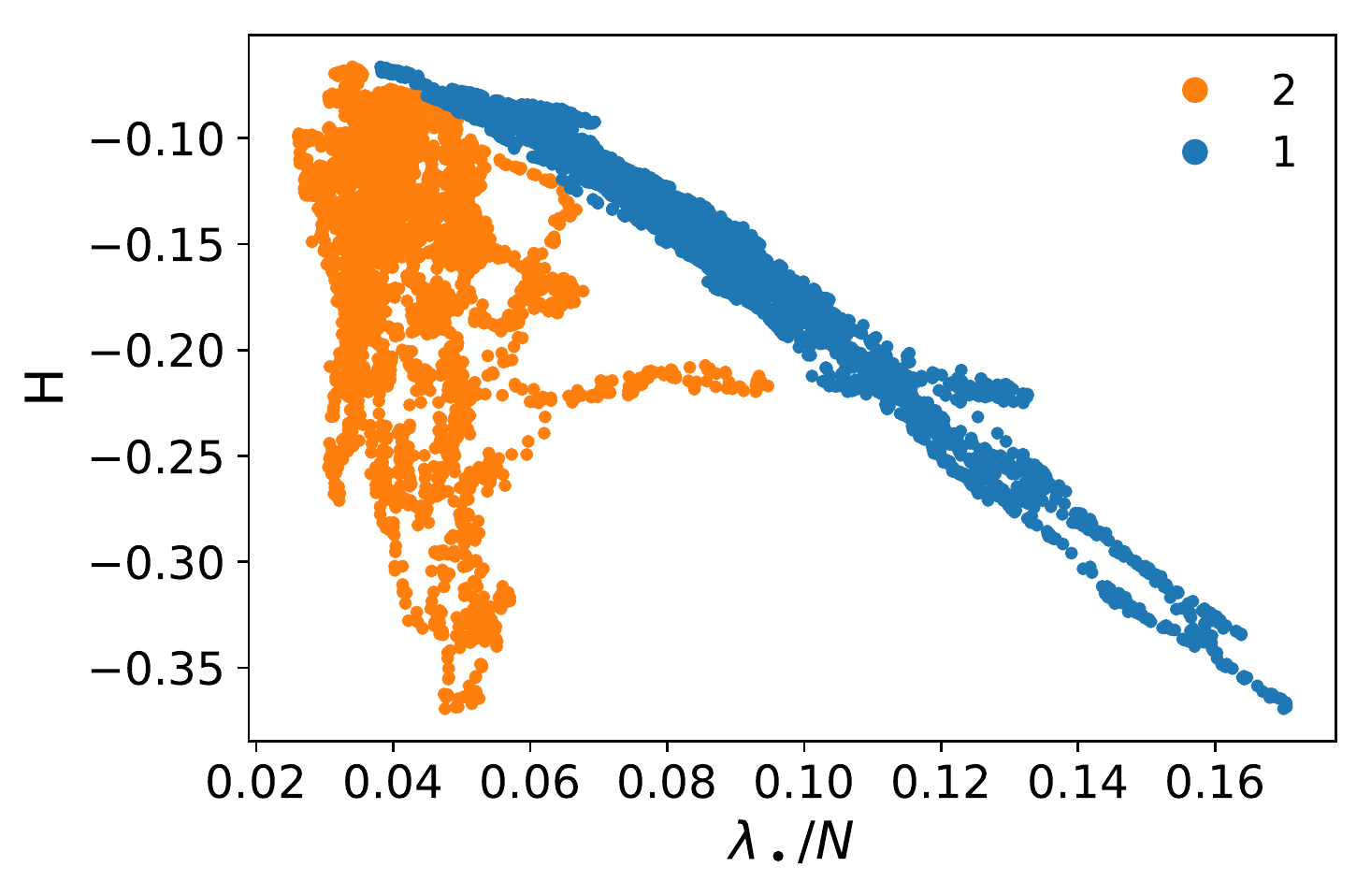}}
\subfigure[\label{fig:rhov}]{\includegraphics[width=0.4\textwidth]{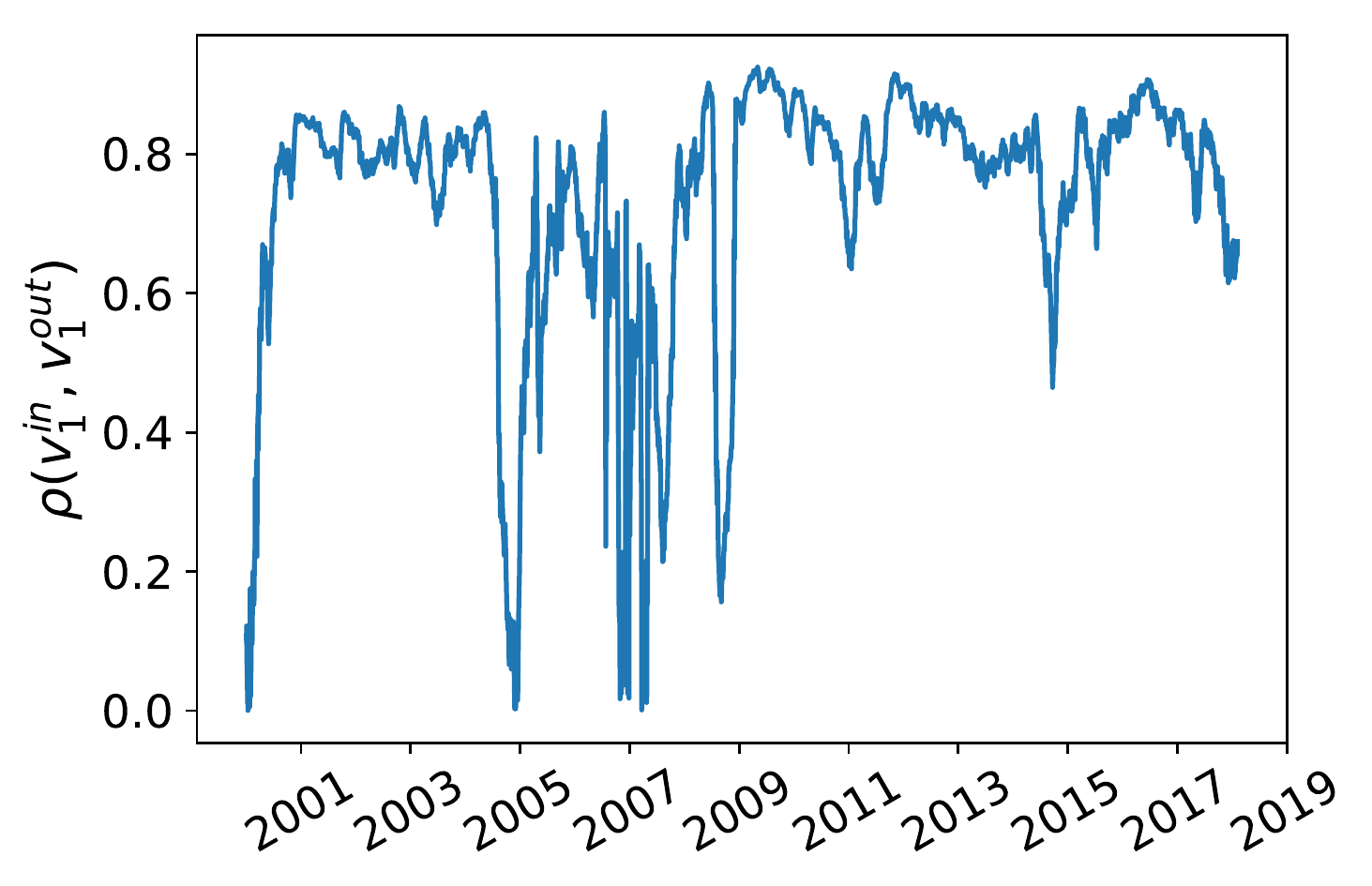}}
\caption{\small{$(a)$ Scatter plot of $H$ as a function of the fraction of variance explained by the largest and the second largest  eigenvalues; $(b)$ Pearson correlation between the eigenvector corresponding to the largest eigenvalue in the in-sample time window $\textbf{v}_1^{in}$ and in the out-of-sample time window $\textbf{v}_1^{out}$. $T_{in}=T_{out}=100$}}
\end{figure}

The components of the first eigenvector of the correlation matrix $C$ (of real returns $r$) typically have the same sign because the 
market mode is still present; however, since we are interested in the macroscopic cluster composition of the stocks, we removed the  market  mode defined as the median return when computing $\Phi$. This means that the largest eigenvector $\mathbf{v}_1$ of $\Phi$ will be composed of positive and negative entries, and it will be strongly correlated with the second largest eigenvector of the correlation matrix $C$. As was pointed out by Buccheri et al\cite{buccheri2013evolution}, the dynamics of the second eigenvalue of $C$ is quite independent from that of the first one, and the direction of the related eigenvector is quite stable over time. In fact, in Fig.~\ref{fig:lmdH}, we show that the fraction of variance explained by the largest eigenvalue of $\Phi$, $\lambda_1/N$,  is strongly anti-correlated with $H$ ($R^2=0.94$), whereas  $\lambda_2/N$ is only weakly anti-correlated with $H$. Furthermore, we show in Fig.~\ref{fig:rhov} that the direction of $\mathbf{v}_1$ is quite stable across the time periods. 

The main implication of this observation is that, if the largest eigenvalue $\lambda_1$ ($\Phi$)  increases and if the direction of $\textbf{v}_1$ does not change substantially, then the stability of some the triads increases. This is when one can predict the sign of some correlations according to Heider's balance theory. 

\subsection*{Prediction of correlation signs}\label{sec:predict}
As observed in the previous section, triads suggest a mechanism to predict the future states of $S$, i.e., of the signs of correlation coefficients. For this purpose we define $\Delta_{ij}$ as the contribution of asset pair $(i,j)$ to $H$, which amounts to (in matrix notation)
\begin{equation}
\Delta = \frac{S \circ S^2}{N-2},
\end{equation}
where  $S$ is the signed adjacency matrix defined above and $\circ$ is the Hadamard (element-wise) product. Note that $\Delta_{ij}\in[-1,+1]$: $\Delta_{ij}=+1$ if the link $(i,j)$ forms stable pairs with all the other nodes, and $\Delta_{ij}=-1$ if the link $(i,j)$ forms unstable pairs with all the other nodes.

Our hypothesis is that the lower $\Delta_{ij}$ in the in-sample window, the higher the probability for the link $(i,j)$ to switch its sign in the future, and reversely: high values of $\Delta_{ij}$ should be related to high out-of-sample stability interactions. 

In order to test this hypothesis, we build a binary classifier that uses $\Delta_{ij}$ as discrimination variable. Specifically, we evaluate $\Delta_{ij}$ in an in-sample time window of $T_{in}$ days, and we try to predict the sign switch of $\Phi_{ij}$ in the out-of-sample of $T_{out}$. In our experiments the in-sample and out-of-sample time windows are not overlapping and in general $T_{in} \neq T_{out}$.  In order to assess the ability of $\Delta_{ij}$ to predict the sign stability, we used the Receiver Operating Characteristic (ROC) curve\cite{hanley1982meaningroc}, a graphical representation of the True Positive Rate as a function of the False Positive rate as the discrimination threshold varies.  As a summary of the performance of a discrimination variable, we use the Area Under the Curve (AUC). We therefore compare the ROC curves obtained with $\Delta_{ij}$ as discrimination variable, and the other associated with the value of the correlation $\Phi_{ij}$. 

\begin{figure}
\centering
\subfigure[\label{fig:roc}]{\includegraphics[width=0.4\textwidth]{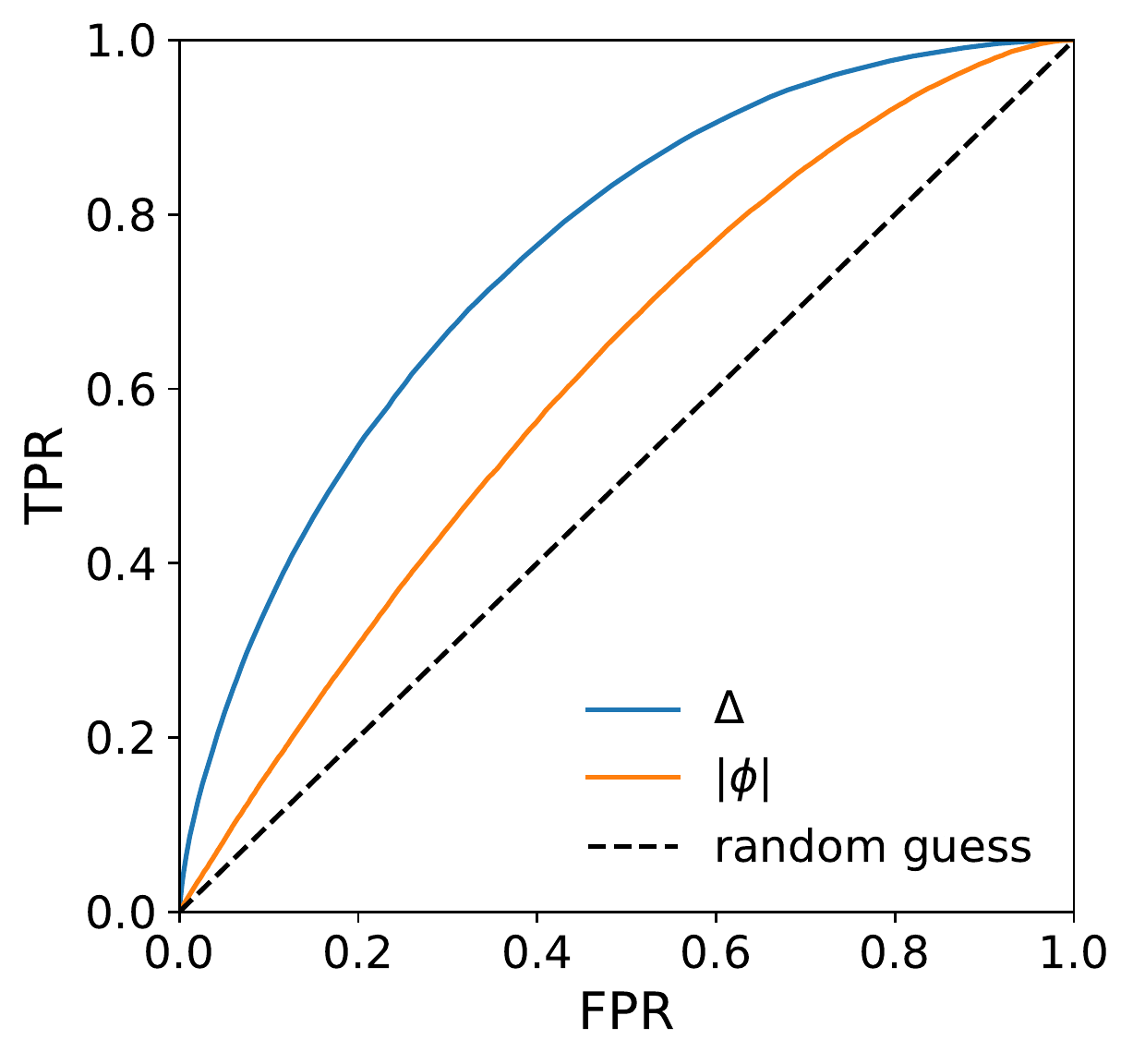} }
\subfigure[\label{fig:pstab}]{\includegraphics[width=0.5\textwidth]{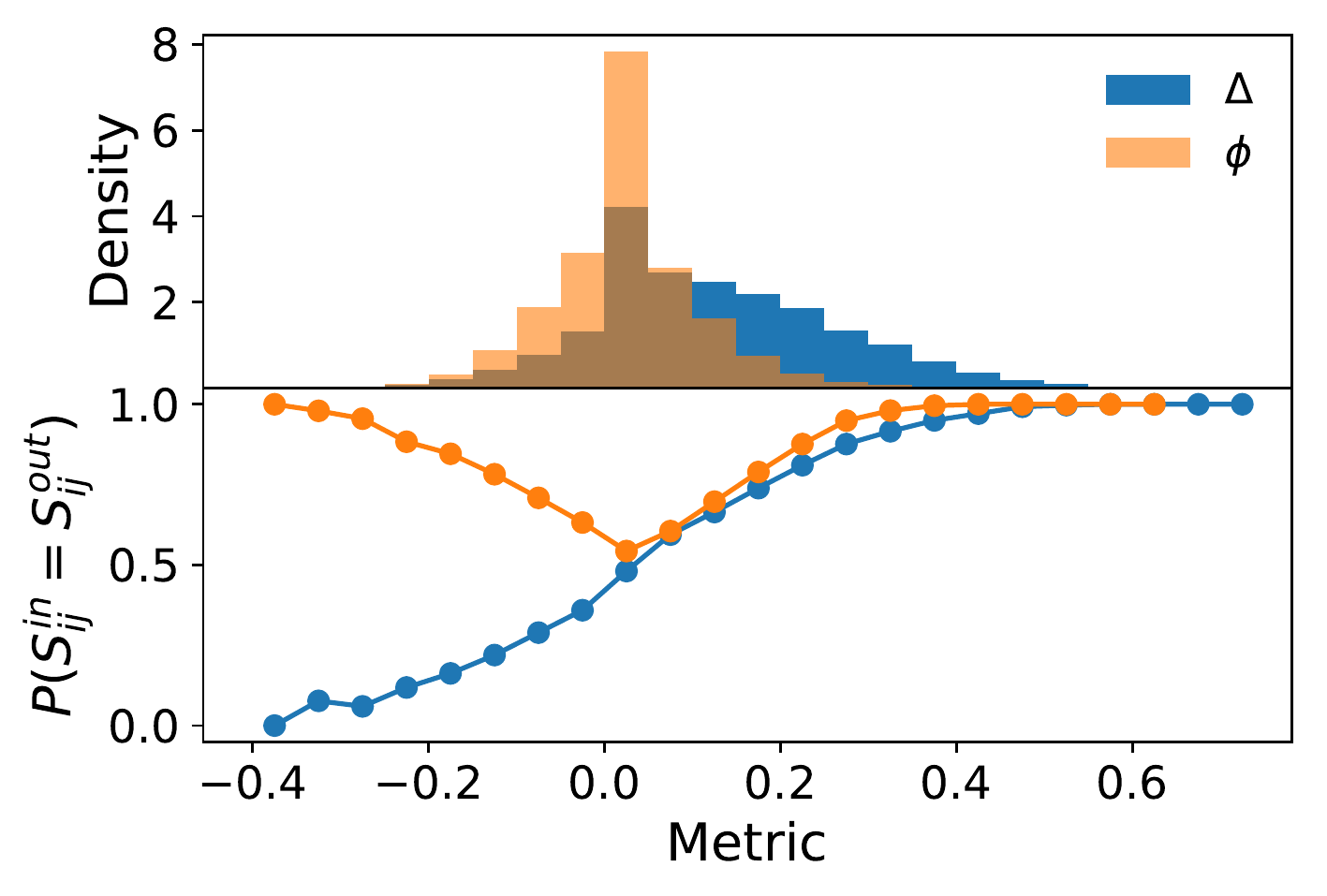} }

\caption{\small{$(a)$ ROC curve for 2011-18-04 with $T_{in}=T_{out}=155$; $(b)$ the lower subplot is the probability to preserve the in-sample sign in the out-of-sample on 2011-18-04 for different values of the discrimination parameter binned in steps of $0.05$, the upper subplot is the related marginal distribution.}}
\end{figure}

Intuitively, larger correlations (in absolute value) should be more stable than  smaller ones, if only because of estimation noise. In the high-dimensional case however, stability depends more on triadic relationships that on the intensity of correlations.
In Fig.~\ref{fig:roc}, we show an example of the ROC evaluated on 2011-18-04 with $T_{in}=T_{out}=155$. The variable $\Delta_{ij}$ outperforms $|\Phi_{ij}|$ for the prediction of the correlation sign: their respective AUCs are $0.75$ and $0.61$. The origin of this difference clearly appears in Fig.~\ref{fig:pstab} which plots the probability that both in- and out-of-sample signs are the equal as a function of the discrimination parameter. The first obvious observation is that $|\Phi|$ is not able to predict the sign changes as $P(S_{ij}^{in}=S_{ij}^{out})$ is never smaller than $0.5$. Furthermore, by looking at the marginal distributions, it is clear that most of the correlations lie close to $|\Phi|=0$; this explains why $|\Phi|$ is only slightly more informative than a coin toss. On the other hand, the marginal distribution of $\Delta$ has better coverage, resulting in a better correction sign prediction performance. This result also holds for the correlation coefficients of raw returns $C$.

\begin{figure}
\centering
\subfigure[\label{fig:aucdelta}]{\includegraphics[width=0.32\textwidth]{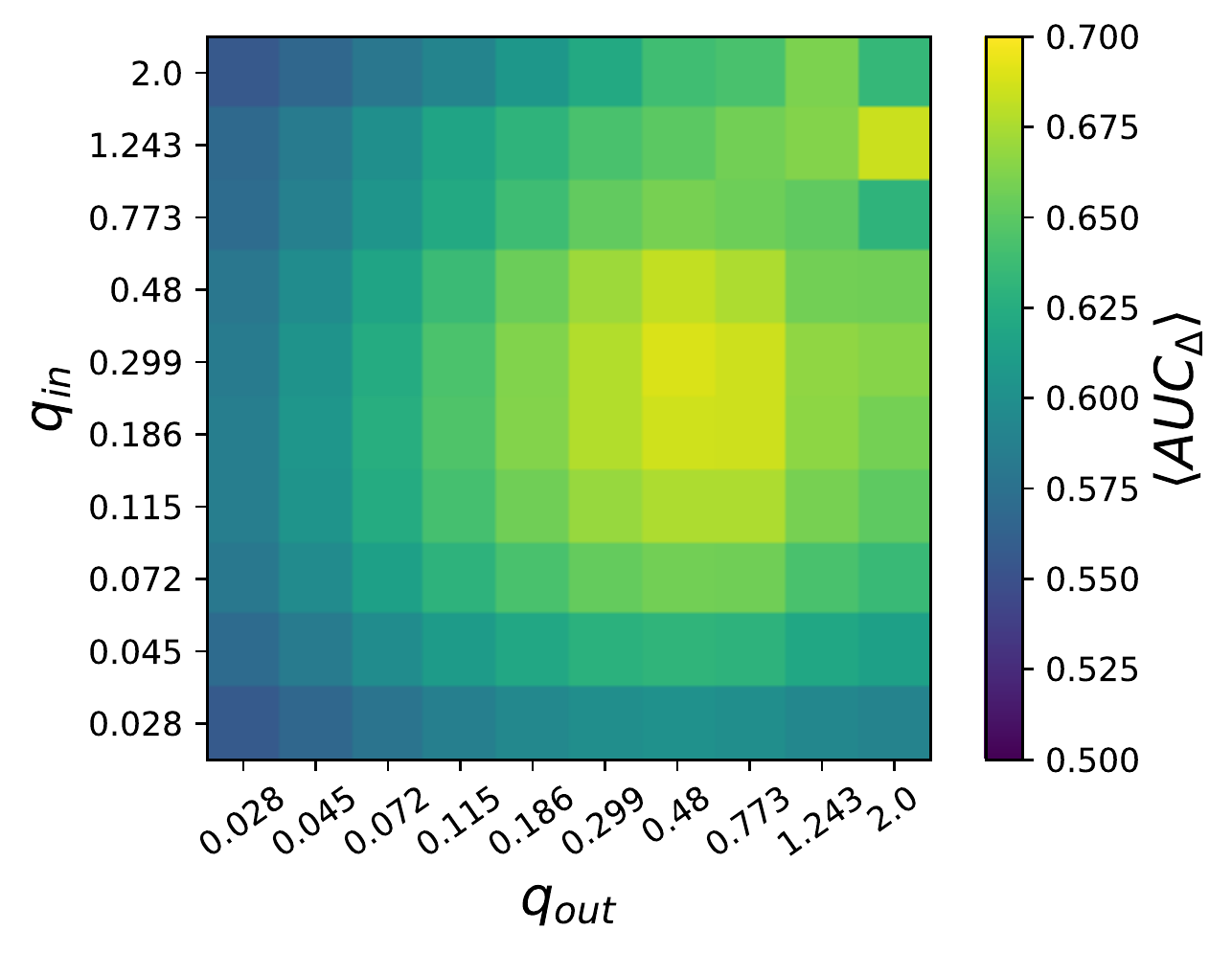}}
\subfigure[\label{fig:aucphi}]{\includegraphics[width=0.32\textwidth]{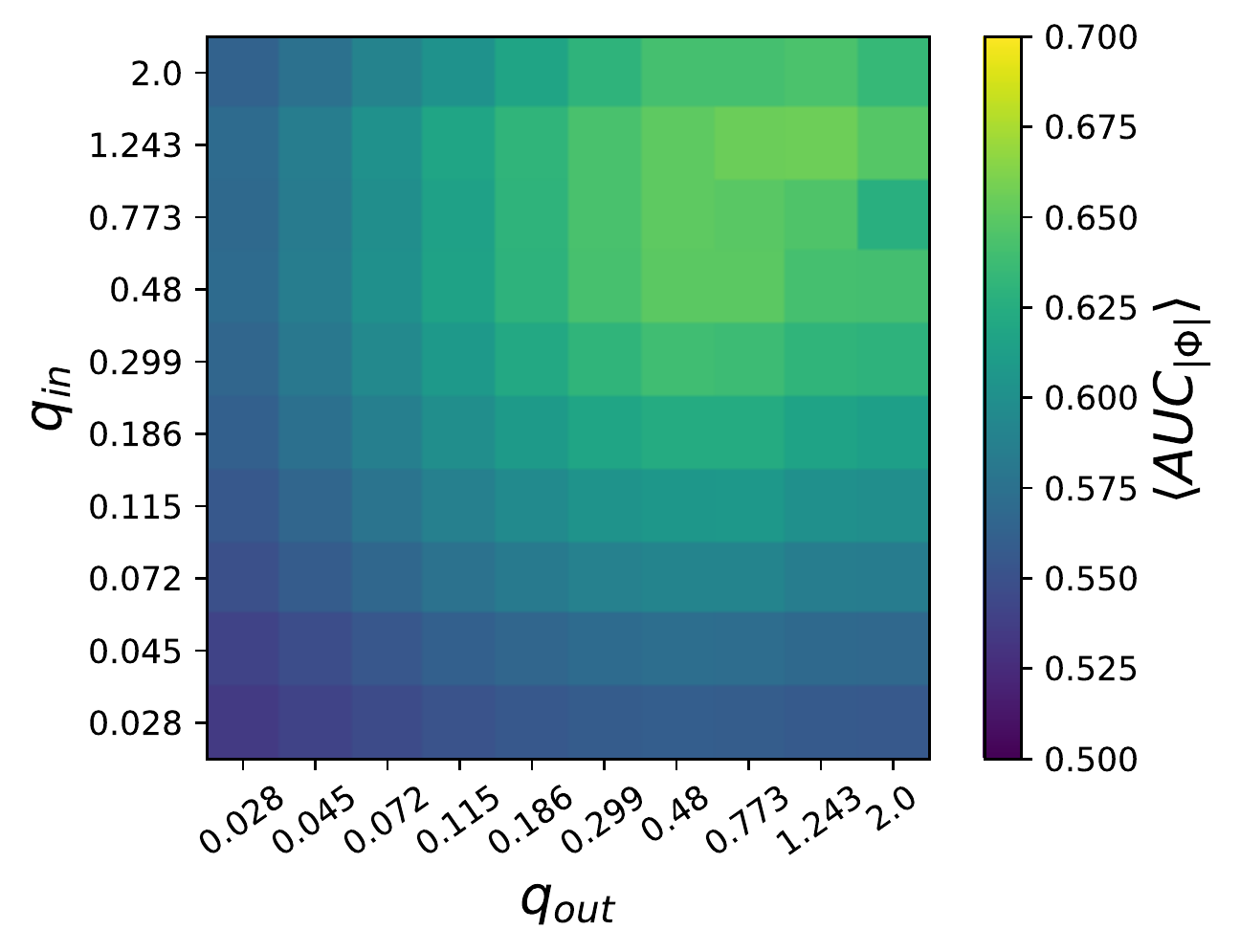}}
\subfigure[\label{fig:aucdiff}]{\includegraphics[width=0.32\textwidth]{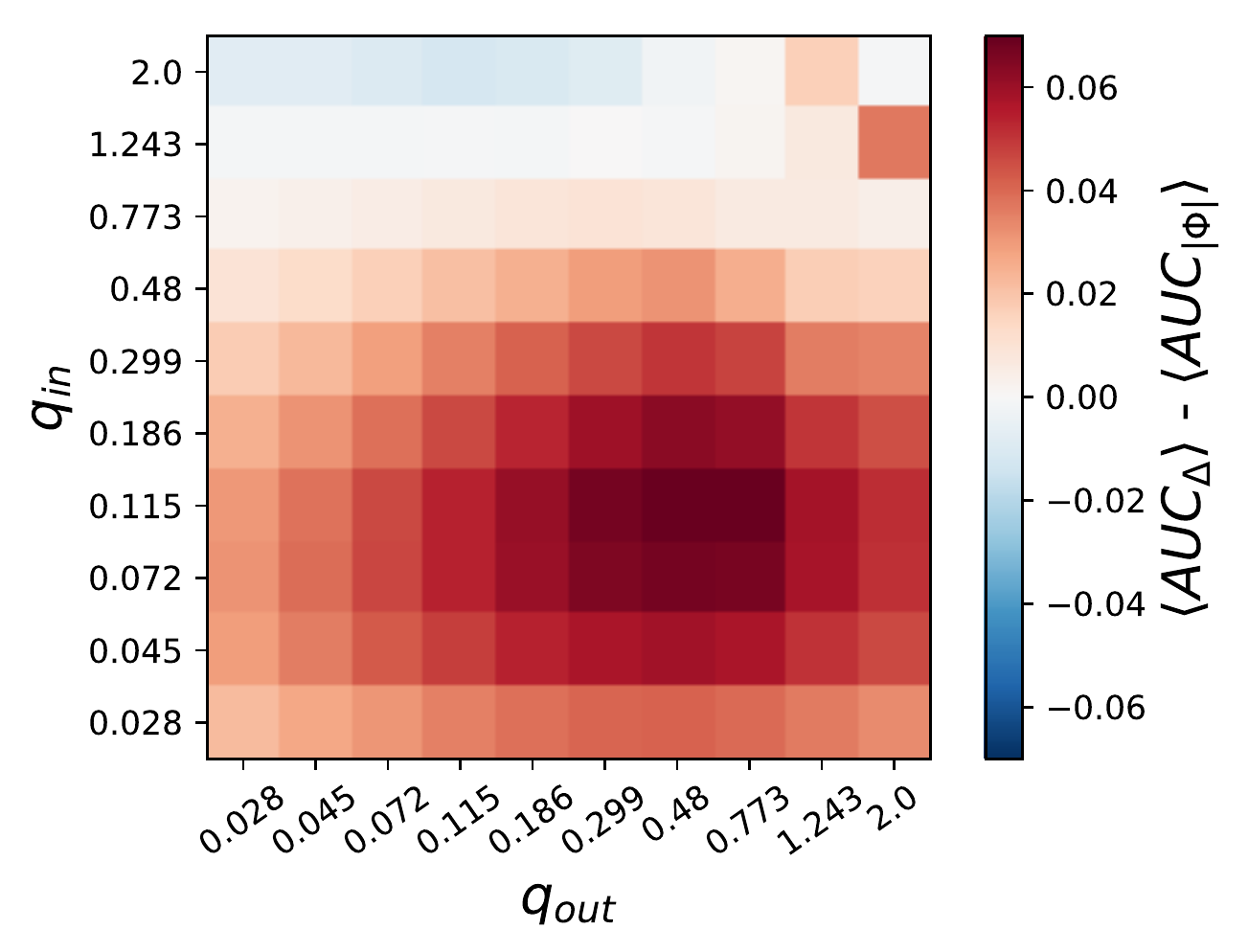} }

\caption{\small{$(a)$ Heatmap of the difference between the average AUC of the two discrimination variables $\Delta$ and $|\Phi|$; $(b)$ heatmap of the average AUC$_{\Delta}$; $(c)$ heatmap of the average AUC$_|\Phi|$. }}
\end{figure}

We computed the performance of both predictors for a wide range of calibration and test window lengths chosen in order to include partial-rank and full-rank correlation matrices, i.e., 10 values between 20 and 2000 with a geometric progression. For each pair $(T_{in},T_{out})$, we estimate the AUC of each method in rolling windows with a step of 1 day. In Fig.~\ref{fig:aucdelta} and \ref{fig:aucphi}, we show the average AUC in considered time-period for predictors $\Delta$ and $|\Phi|$ respectively. To highlight the different behavior, Figure~\ref{fig:aucdiff} shows the difference $\langle \mbox{AUC}_{\Delta}\rangle - \langle \mbox{AUC}_{|\Phi|}\rangle$, where  $\langle \mbox{AUC}_{X}\rangle$ is the average value of the AUC in the considered time-period for predictor $X\in\{\Delta,|\Phi|\}$.

 The dependence of the AUC as a function of $q_{in}$ and $q_{out}$ is worth discussing: first, $\textrm{AUC}_{|\Phi|}$ increases monotonically as a function of both $q_{in}$ and $q_{out}$; second, $\textrm{AUC}_{\Delta}$ has a local maximum at about $(q_{in},q_{out})\simeq (0.1,0.5)$, i.e., deep in the high-dimensional regime. The difference between the two is clear: triads are better as long as $q>1$ and correlations better when $q<1$. The same results hold for Hong Kong equities data (see SI).

\begin{figure}
\centering
\subfigure[\label{fig:auccomp}]{\includegraphics[width=0.4\textwidth]{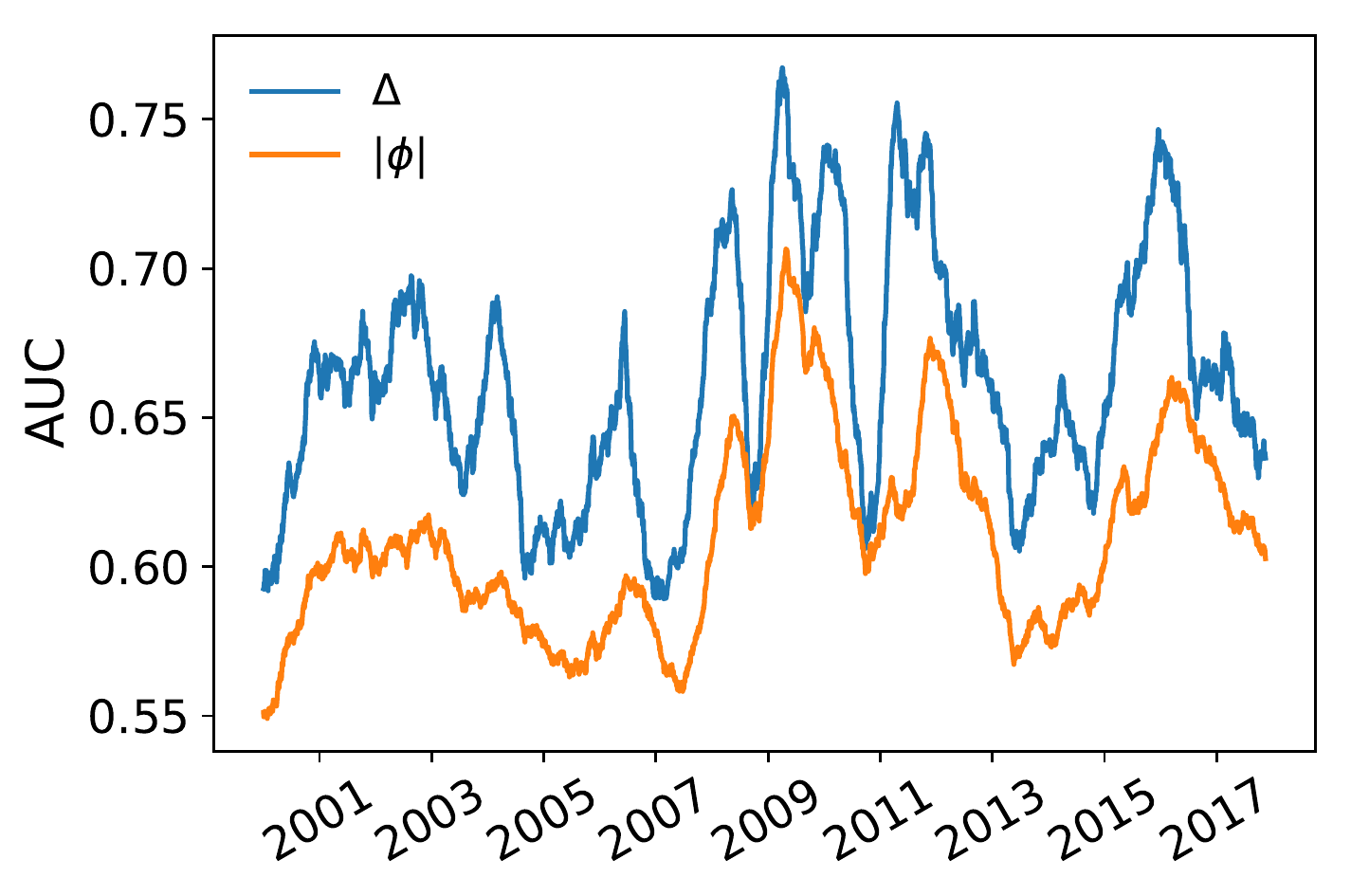}  }
\subfigure[\label{fig:aucH}]{\includegraphics[width=0.46\textwidth]{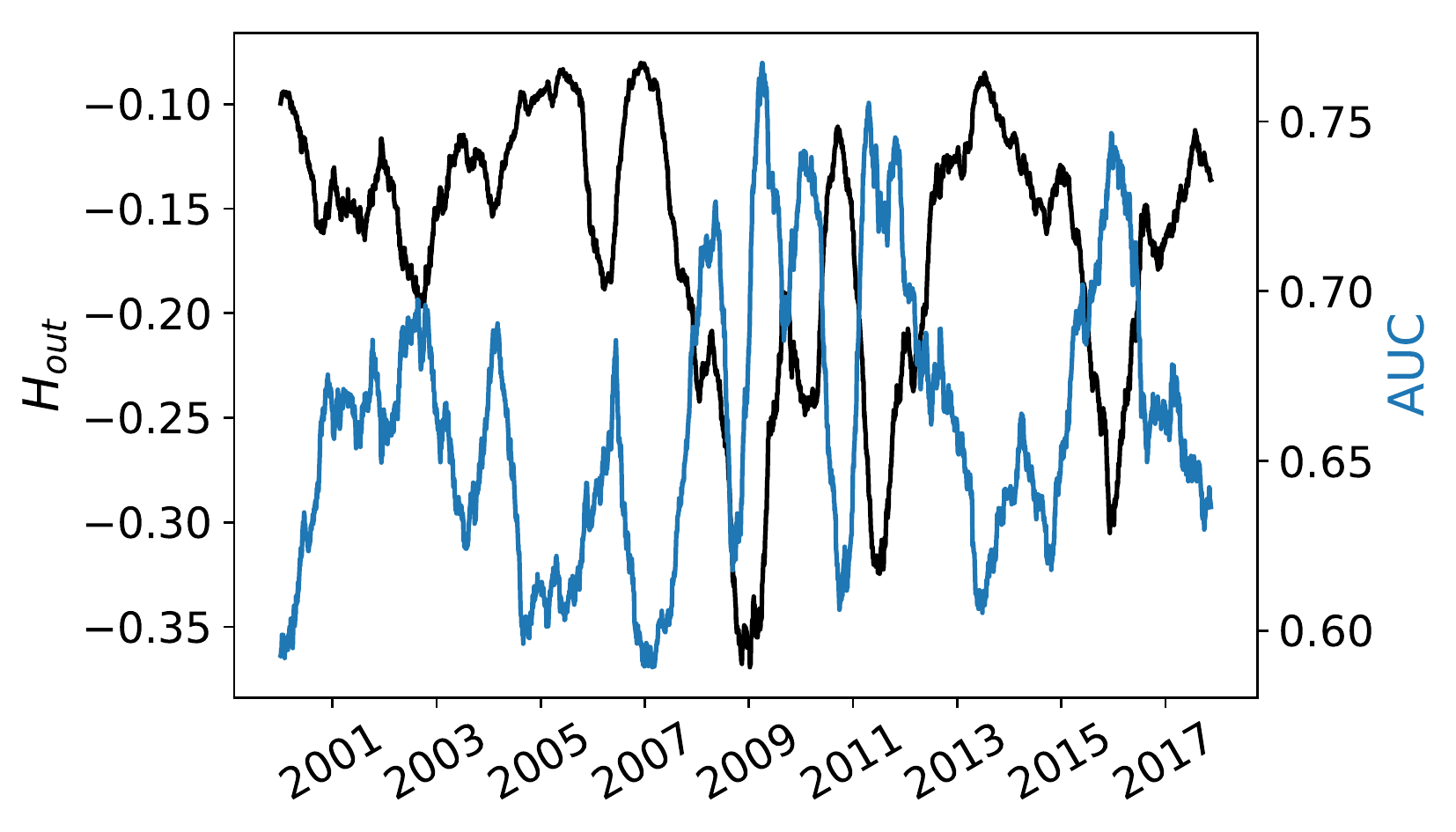} }
\caption{\small{$(a)$ Evolution of AUC for the two model; $(b)$  AUC (blue line) and H evaluated on the out-of-sample as a function of time. Both panels refer are evaluated with $T_{in}=T_{out}=155$}}
\end{figure}

Figure~\ref{fig:auccomp} shows the evolution of the AUC for $T_{in}= T_{out}=155$. Although  $\Delta$ outperforms $|\Phi|$ most of the time, the difference between the methods is not constant. On 2008-26-06 for example, both AUCs are almost equal. Fig.~\ref{fig:aucH} illustrates the strong anti-correlation between AUC$_\Delta$ and the out-of-sample $H$. Specifically, the two variables have a Pearson correlation of $-0.77$ and a Spearman correlation of $-0.83$, with a p-value close to $0$. In fact, we must consider that when $H$ increases,  the total number of stable pairs decrease, and reversely. In any case, even in the worse situation, the variable $\Delta$ performs as well as $\Phi$.

\section*{Discussion}\label{sec:discuss}

In the high-dimensional regime, correlation matrices become pathologically noisy and the strength of their coefficients are not the best predictors of their stability. Accounting for more complex relationships between correlations makes it possible to predict the sign change of correlation coefficients deep in this regime. More precisely, dyadic relationships are better predicted from triadic relationships, as higher-order nonparametric structures exploit non-obvious structure of high-dimensional correlation matrices.

Potential applications of our method includes building better portfolios by accounting for predicted correlation sign changes, which will be addressed in a future work.

\section*{Methods}

\subsection*{Statistically Validated Networks}\label{sec:SVN}
A binarized version of the return matrix $\textbf{b} \in N \times T$ can be interpreted as a bipartite network. A bipartite network is a particular network where the nodes belong to two different sets, in our case one set is composed by the $N$ stocks and the other set by the $T$ days. Only links among nodes of different sets are allowed. Specifically, a stock is linked to a day if its return is larger than the median return of all the stocks in such day. A typical approach to study bipartite networks is to project them into monopartite networks. A projected network is a network composed by nodes of only one set, and a link among to node is established only if those nodes share at least one common neighbors in the opposite set. However, this linkage rule is too permissive and typically leads to a very dense projected network; therefore, the resulting interaction topology could be in many cases meaningless. An alternative approach, defined in Tumminello et al\cite{tumminello2011statistically}, is to link two nodes of the same set if the number of common neighbours they share in the opposite set cannot be explained by random chance. Specifically, one computes a p-value for each link according to the cumulative hypergeometric distribution:
\begin{equation}
\pi_{ij} = 1-\sum_{x=0}^{c_{ij}-1}  \frac{{k_j \choose k_i}{{T-k_j} \choose {x-k_i}}}{{n \choose x}},
\end{equation}
where $c_{ij}$ is the number of common neighbours of $(i,j)$, $k_i$ and $k_j$ are the degree of the nodes $i$ and $j$ in the bipartite network respectively, and $T$ is the number of nodes of the opposite set. Since the test is performed on every link of the projected network, a multiple-comparison correction is required to control the fraction false positive discoveries; in this work we use the False Discovery Rate (FDR) which guarantees that the proportion of false discovery is strictly less than $\alpha$.

It is worth noticing that the number of common neighbours of two nodes can be evaluated with the scalar product $c_{ij} = \sum_t b_{it} b_{jt}$, and the expected number of common neighbours according with the hypergeometric distribution is $E[c_{ij}] = \frac{k_i k_j}{T} = \frac{\sum_t b_{it}\sum_t b_{jt}}{T}$. Therefore, the condition behind the statistical test can be translated into a condition of positivity of the correlation coefficient $\phi_{ij}$.

\section*{Acknowledgements}

This publication stems from a partnership between CentraleSup\'elec and BNP Paribas.

\bibliographystyle{unsrt}
\bibliography{arxive_nonpar}

\section*{Supplementary Information}
\subsection*{Pearson correlations}\label{sec:pearson}
\begin{figure}
\centering
\subfigure[\label{fig:auccompPear}]{\includegraphics[width=0.4\textwidth]{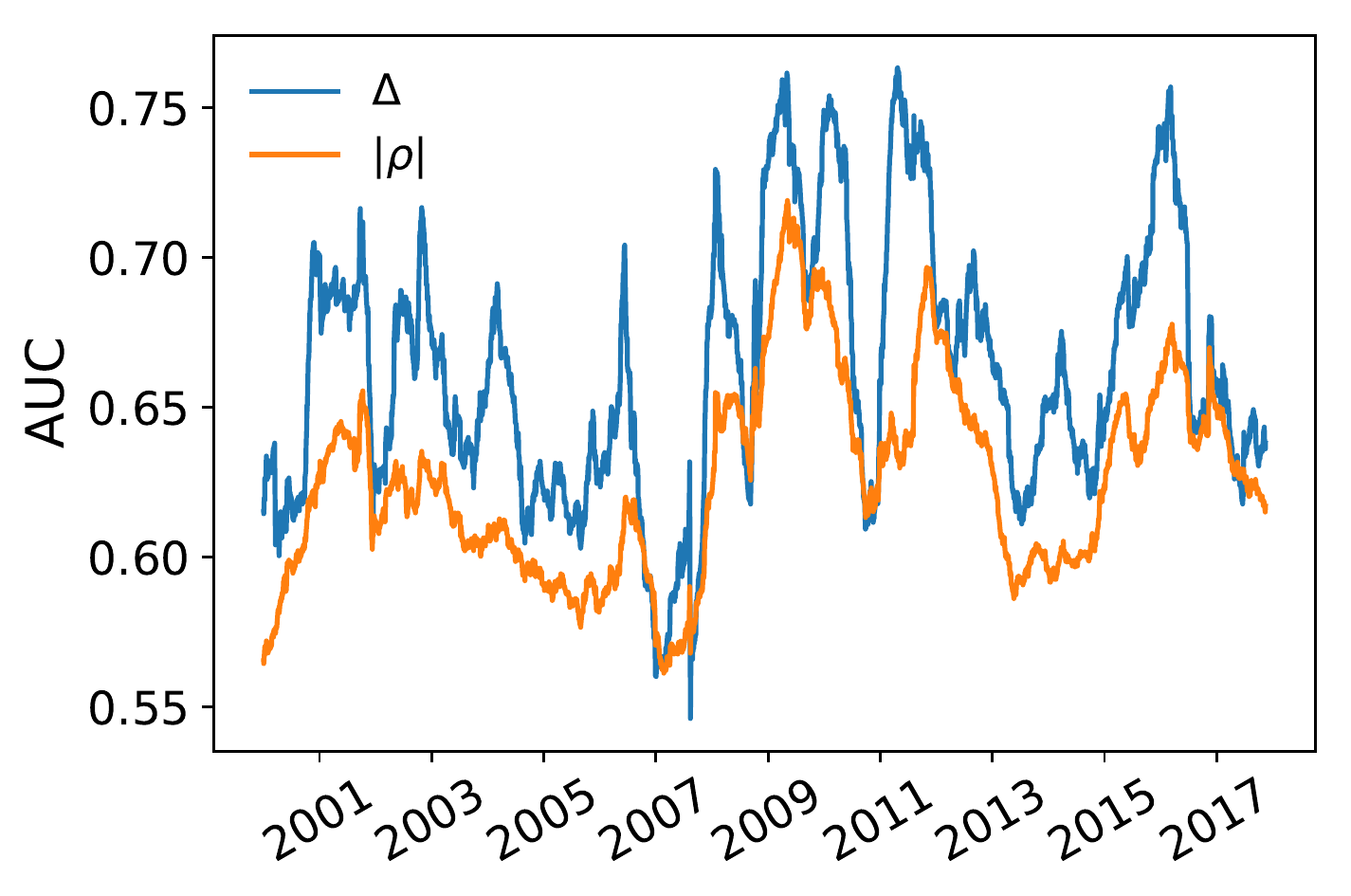}  }
\subfigure[\label{fig:aucDiffPear}]{\includegraphics[width=0.4\textwidth]{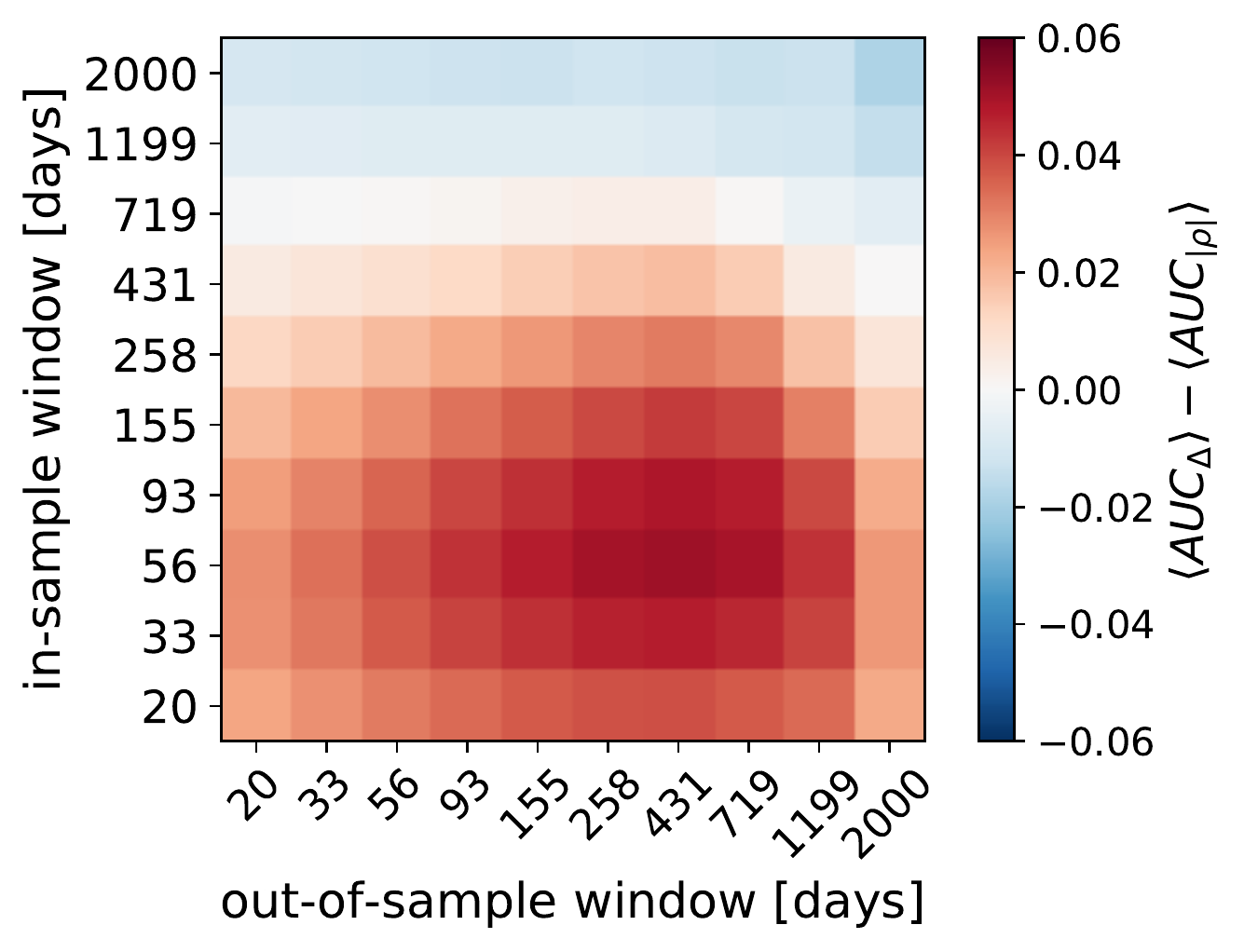} }
\caption{\small{$(a)$ Evolution of AUC for the two model for $T_{in}=T_{out}=155$; $(b)$ Average difference of the AUC among the two model for different in-sample and out-of-sample time windows; Both panels refer to the Pearson correlation matrix among the returns minus the median $\textbf{z}$ }}\label{fig:pear}
\end{figure}

In order to prove that our results are qualitatively independent from the binarization of the returns, we performed the same analysis  from the sign of the Pearson correlation $\rho_{ij}$ computed from partial log-returns. Figure~\ref{fig:pear} confirms that unstable triads are better predictors of correlation sign changes than correlation values themselves.

\subsection*{Market mode removal: first eigenvalue}\label{sec:partcorr}
\begin{figure}
\centering
\subfigure[\label{fig:auccompPPear}]{\includegraphics[width=0.4\textwidth]{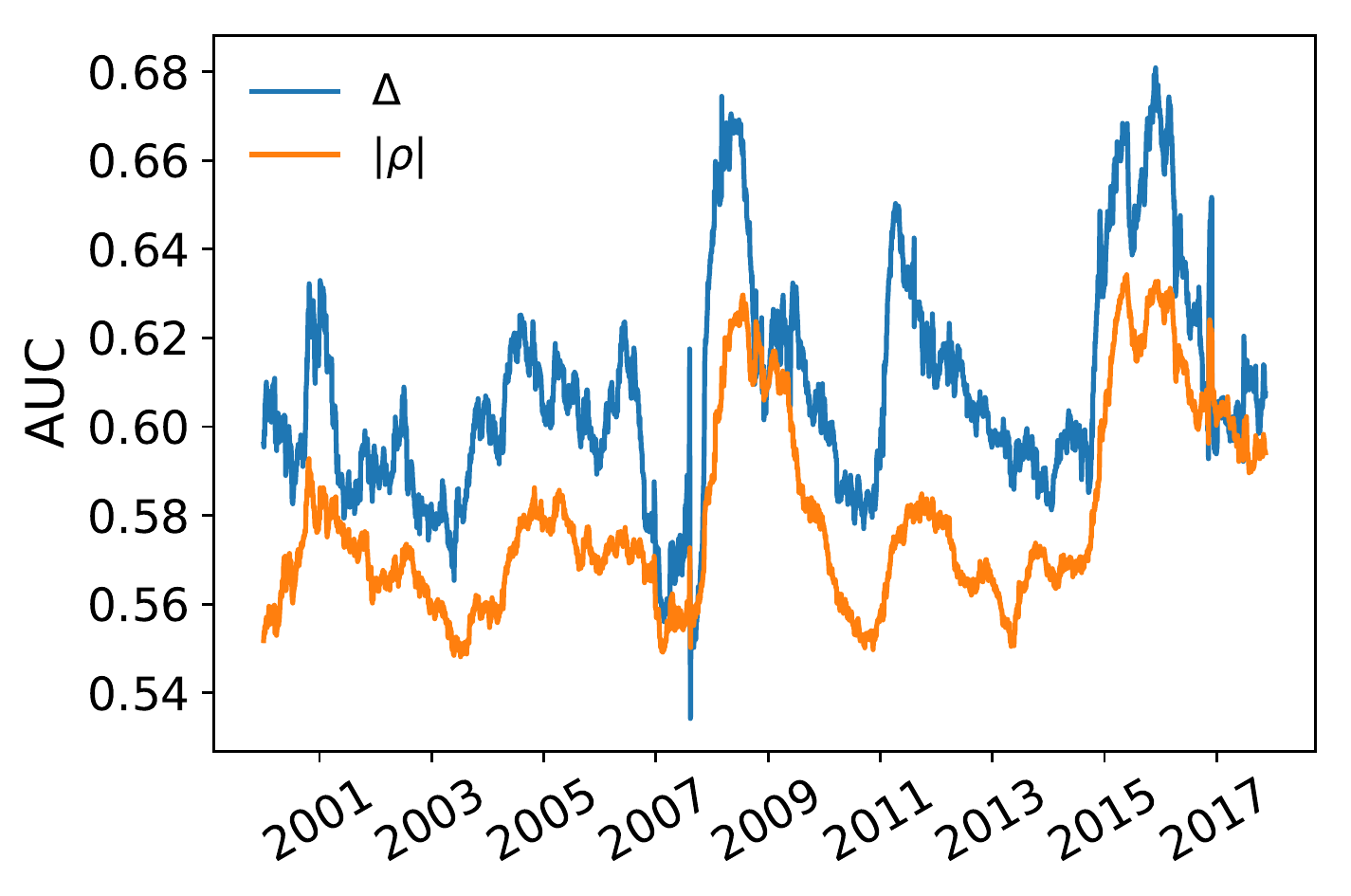}  }
\subfigure[\label{fig:aucDiffPPear}]{\includegraphics[width=0.4\textwidth]{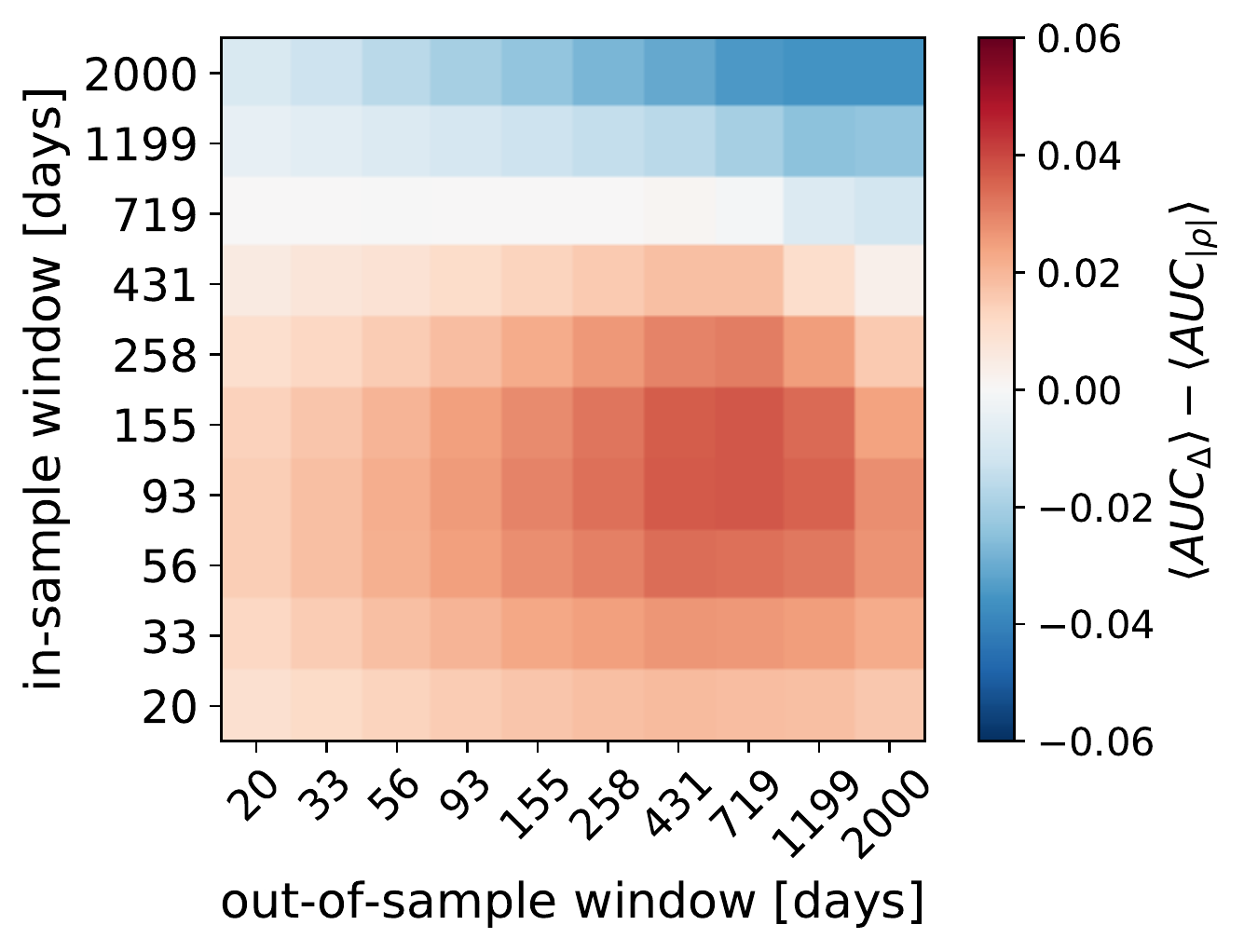} }
\caption{\small{$(a)$ Evolution of AUC for the two model for $T_{in}=T_{out}=155$; $(b)$ Average difference of the AUC among the two model for different in-sample and out-of-sample time windows; Both panels refer to the partial Pearson correlation matrix $\mbox{\boldmath$\rho$}^{(p)}$ among the returns minus the median $\textbf{r}$ }}\label{fig:ppear}
\end{figure}

We show that our results are qualitatively stable if we remove the market mode in a different way. In this section we studied the sign of the partial Pearson correlation matrix obtained with the following equation:
\begin{equation}
\mbox{\boldmath$\rho$}^{(p)} = \sum_{i=2}^N \lambda_i \textbf{v}_i'\,\textbf{v}_i
\end{equation}
where the eigenvalues and the eigenvectors are computed on the Pearson correlation matrix $\mbox{\boldmath$\rho$}$ among the original return matrix $\textbf{r}$. 
As depicted in Fig.~\ref{fig:ppear},  results are similar to those obtained by removing the median returns; however, the nonparametric nature of the estimate is lost. 

\subsection*{Hong Kong Stock Exchange}\label{sec:HK}
In this section we repeated the analysis for the Hong Kong stock exchange. We build a binary return matrix $\textbf{b}$ and the related phi matrix $\mbox{\boldmath$\phi$}$ according with the procedure illustrated in section Data Description and Processing.

\begin{figure}
\centering
\subfigure[\label{fig:HassHK}]{\includegraphics[width=0.32\textwidth]{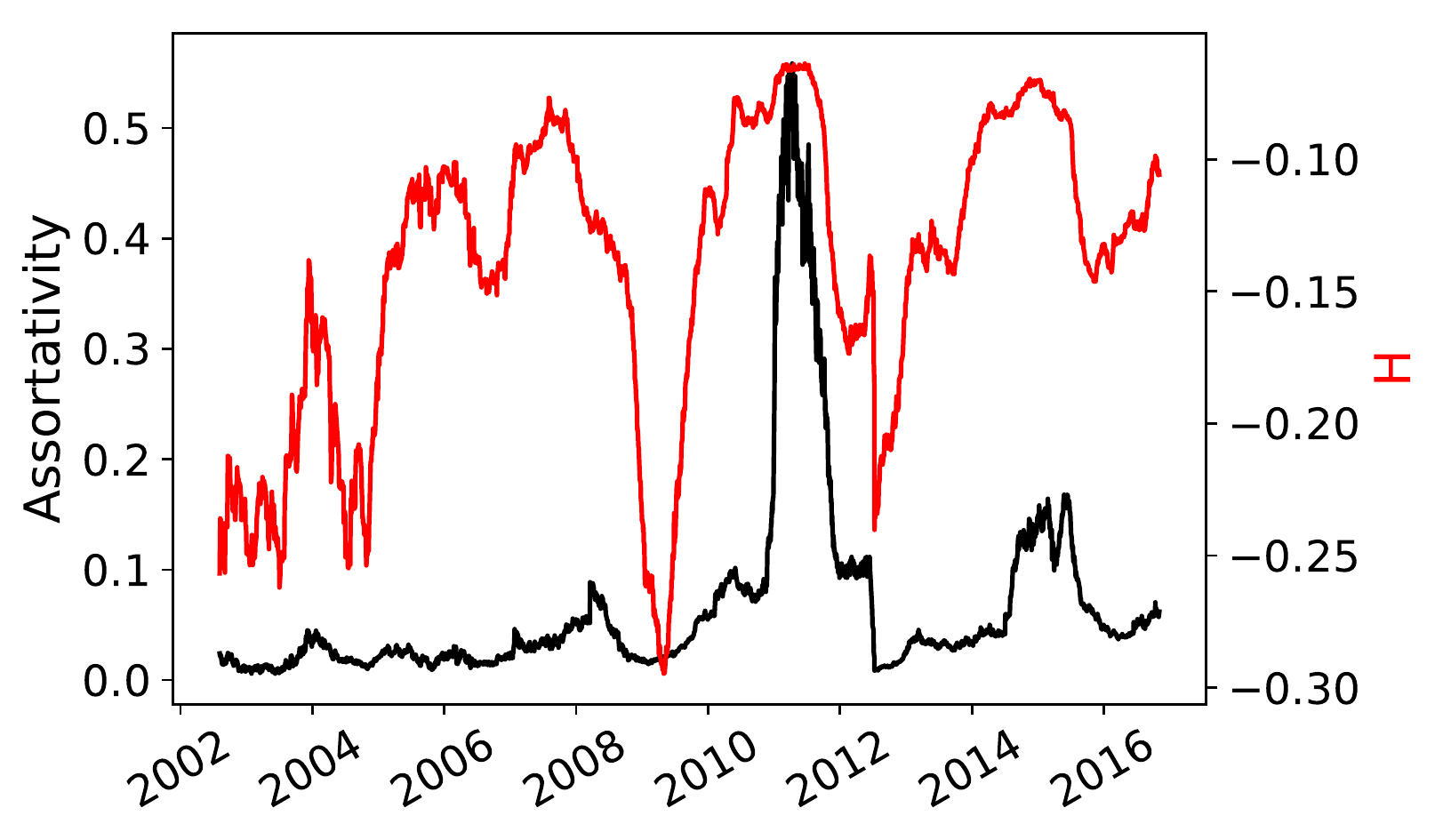}  }
\subfigure[\label{fig:NlinkHK}]{\includegraphics[width=0.32\textwidth]{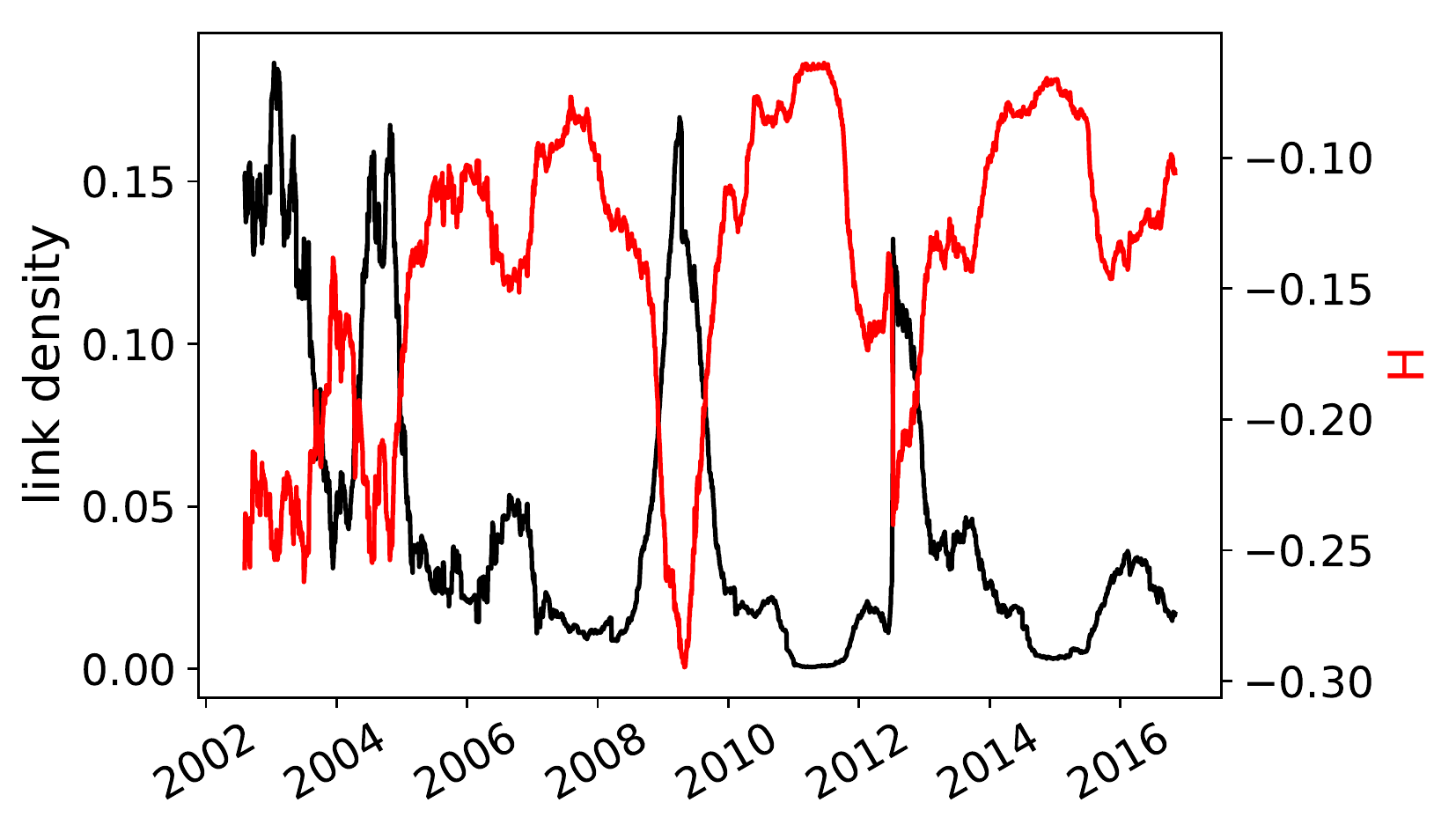}  }
\subfigure[\label{fig:HvolHK}]{\includegraphics[width=0.32\textwidth]{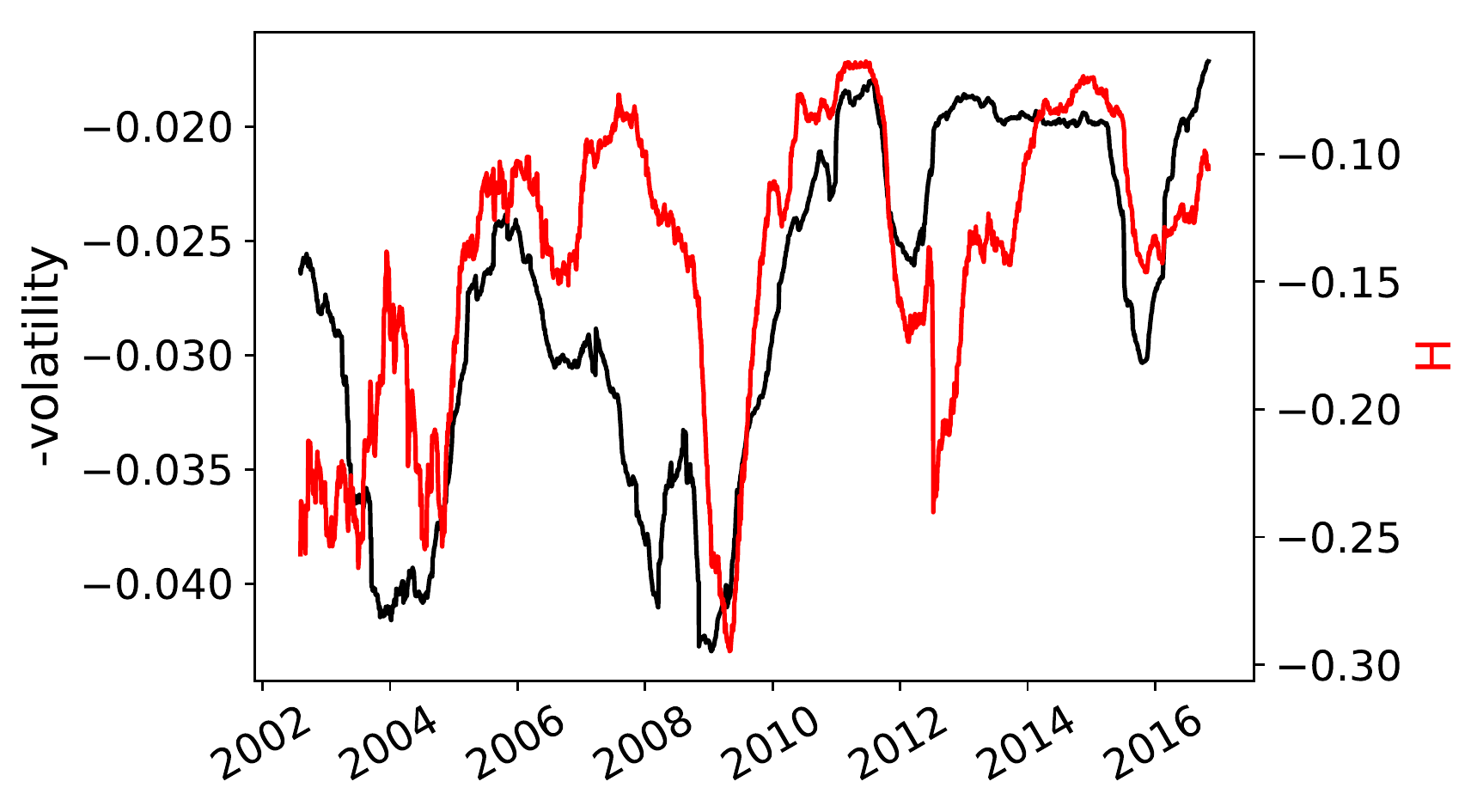} }

\caption{\small{Hong Kong stock exchange: $(a)$ assortativity of the SVN with respect to the sector partition and H; Number links of the SVN and H; $(c)$ minus Volatility and H. $T=100$ days.}}
\end{figure}
In contrast with US equities, we do not observe a strong correlation between the sector partition and the links of the SVN, as shown in Fig.~\ref{fig:HassHK}. In fact, the assortativity is very close to the random null expectation for most of the time-period. We want to stress that this does not necessarily mean that the stocks are not organized in clusters. In fact, as for US equities,  $H$ varies over time, and it is strongly correlated with the number of links of the SVN (Fig.~\ref{fig:NlinkHK}) and with the volatility (Fig.~\ref{fig:HvolHK}). To our knowledge, the sector structure (or apparent lack thereof) of Honk Kong equities has not reported elsewhere.

\begin{figure*}
\centering
\subfigure[\label{fig:aucdiffHK}]{\includegraphics[width=0.32\textwidth]{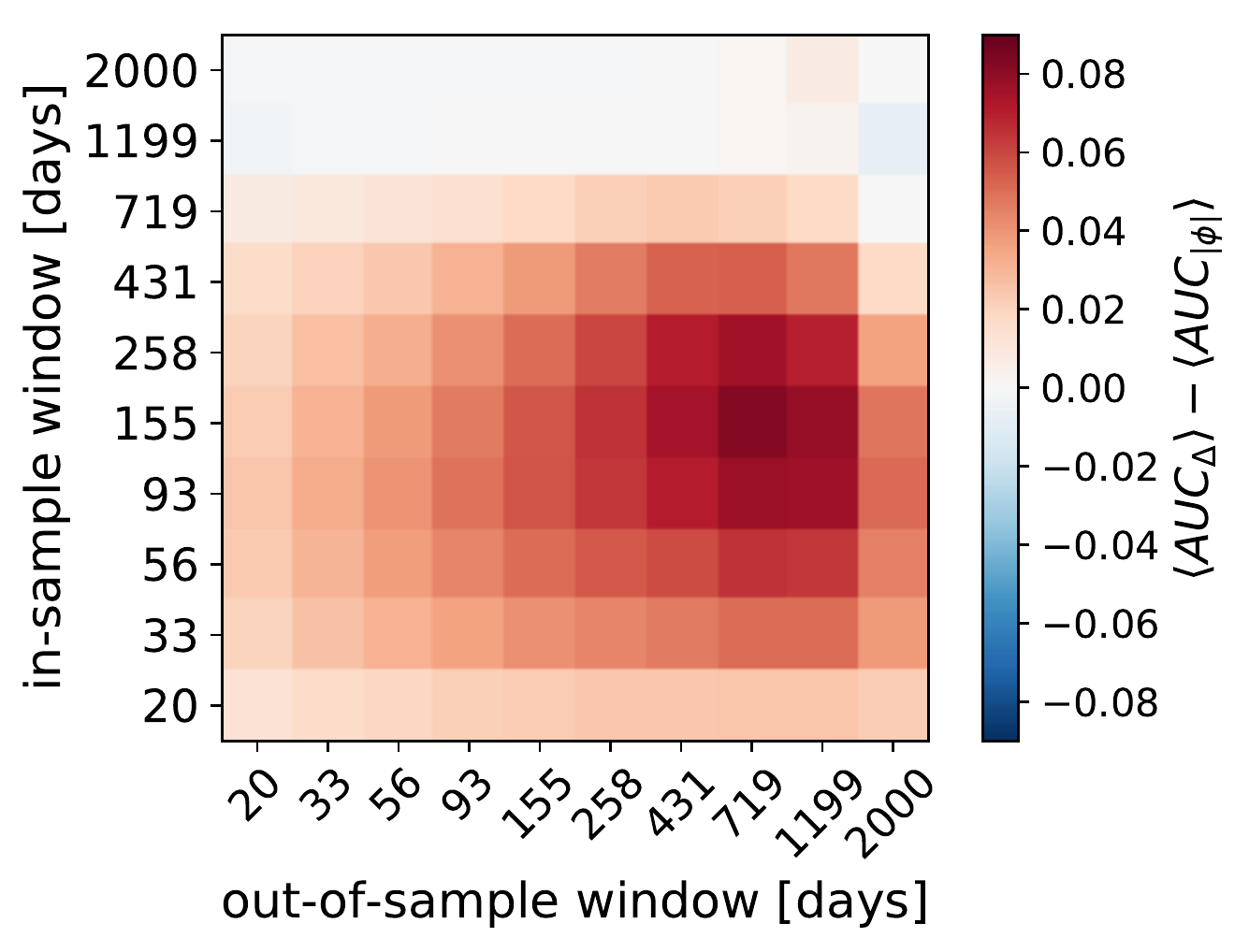} }
\subfigure[\label{fig:auccompHK}]{\includegraphics[width=0.32\textwidth]{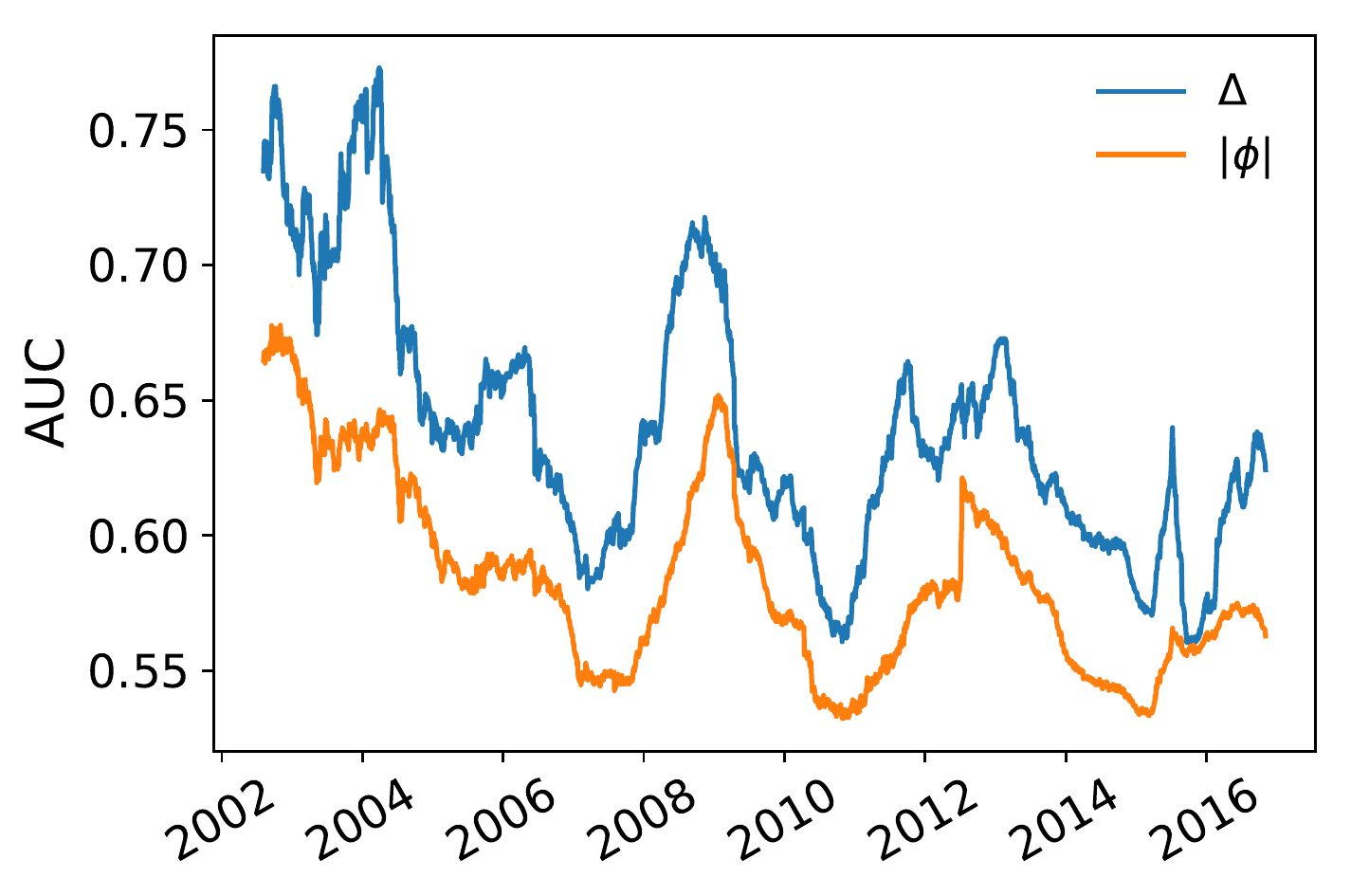}  }
\subfigure[\label{fig:AUCHoutHK}]{\includegraphics[width=0.32\textwidth]{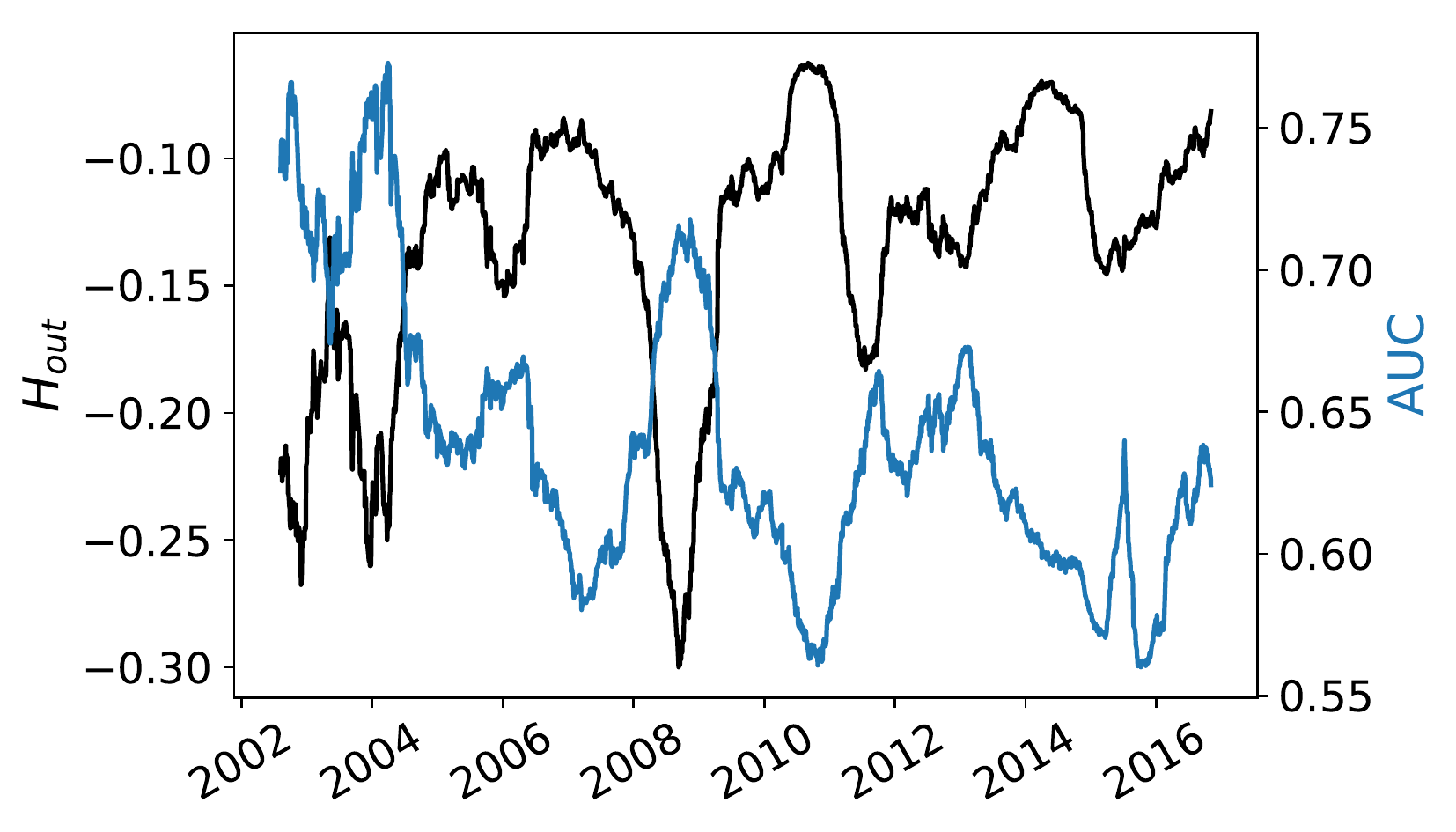} }
\caption{\small{Hong Kong stock exchange: $(a)$ average difference of the AUC among the two model for different in-sample and out-of-sample time windows; $(b)$ Evolution of AUC for the two model for $T_{in}=T_{out}=155$;  $(c)$ AUC and $H$ in the out-of-sample as a function of time for $T_{in}=T_{out}=155$. All panels refer to the Hong Kong Stock Market.}}
\end{figure*}
In this dataset as well, unstable triads are significantly better than the absolute value of the $|\mbox{\boldmath$\phi$}|$ at predicting the instability of correlation signs in the high-dimensional regime (Fig.~\ref{fig:aucdiffHK}).

Once again, dynamical evolution of the AUC (Fig.~\ref{fig:auccompHK}) is minimal in the proximity of a minima of $H_{out}$.

\subsection*{Negative Links}

We report here the SVNs determined between binarized returns of opposite signs. Three observations stand out: i) they are mostly empty; ii) they are more likely to become  non-empty in times of large volatility iii) they are generally disassortative. 

\begin{figure*}
\centering
\subfigure[\label{fig:aucdeltaT}]{\includegraphics[width=0.32\textwidth]{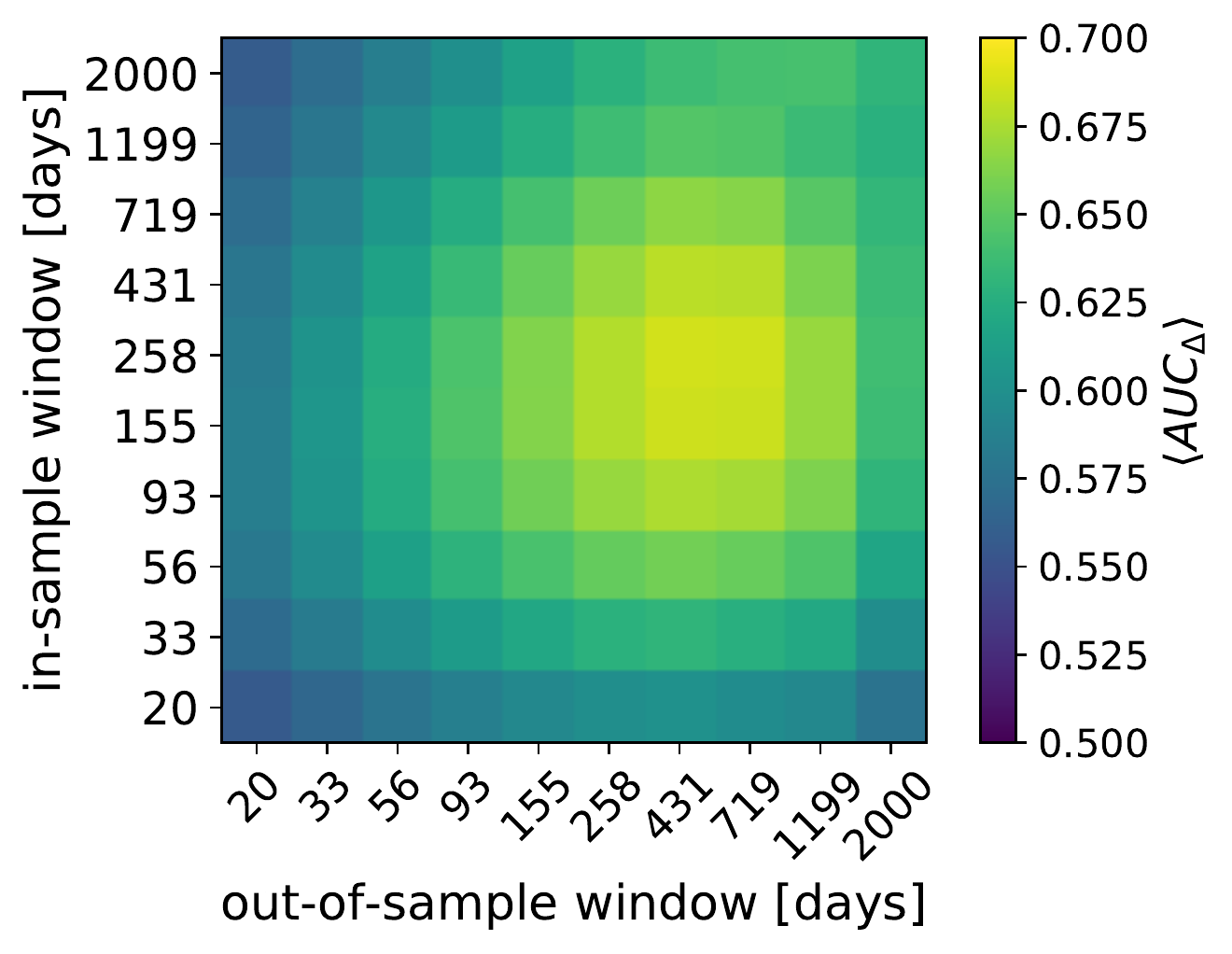}}
\subfigure[\label{fig:aucphiT}]{\includegraphics[width=0.32\textwidth]{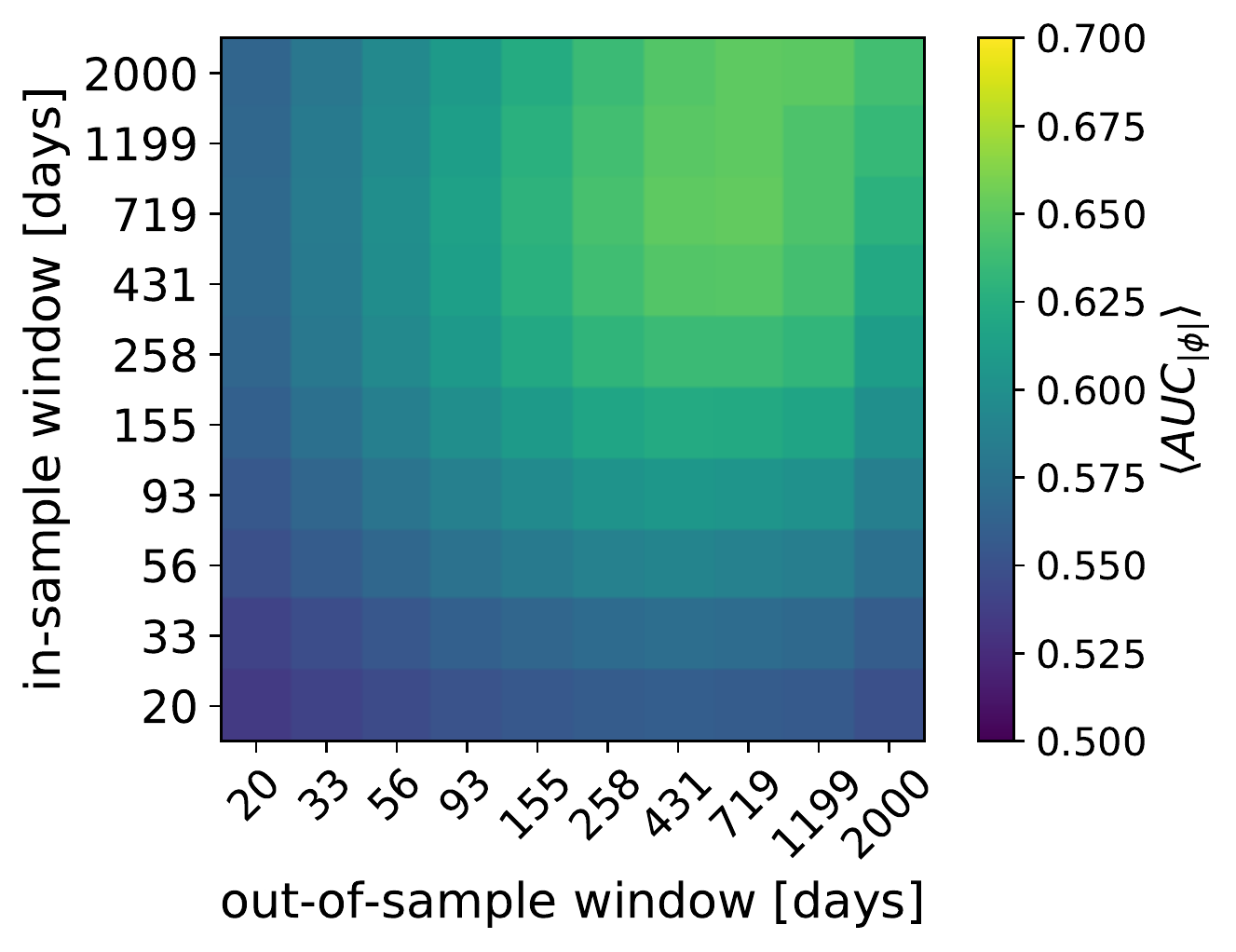}}
\subfigure[\label{fig:aucdiffT}]{\includegraphics[width=0.32\textwidth]{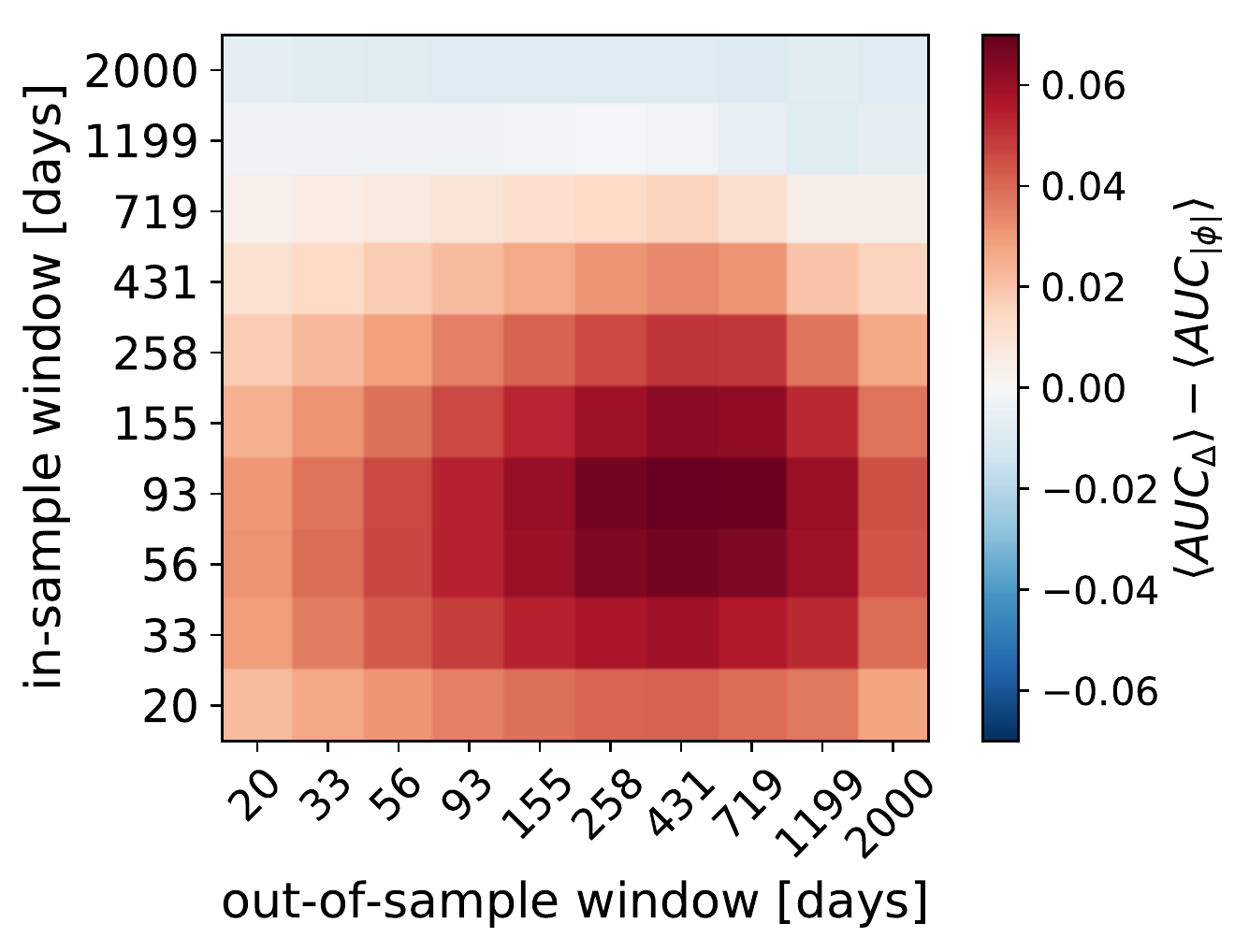} }

\caption{\small{$(a)$ Heatmap of the difference between the average AUC of the two discrimination variables $\Delta$ and $|\phi|$; $(b)$ heatmap of the average AUC$_{\Delta}$; $(c)$ heatmap of the average AUC$_|\phi|$. }}
\end{figure*}

\begin{figure*}
\centering
\subfigure[\label{fig:densUE}]{\includegraphics[width=0.4\textwidth]{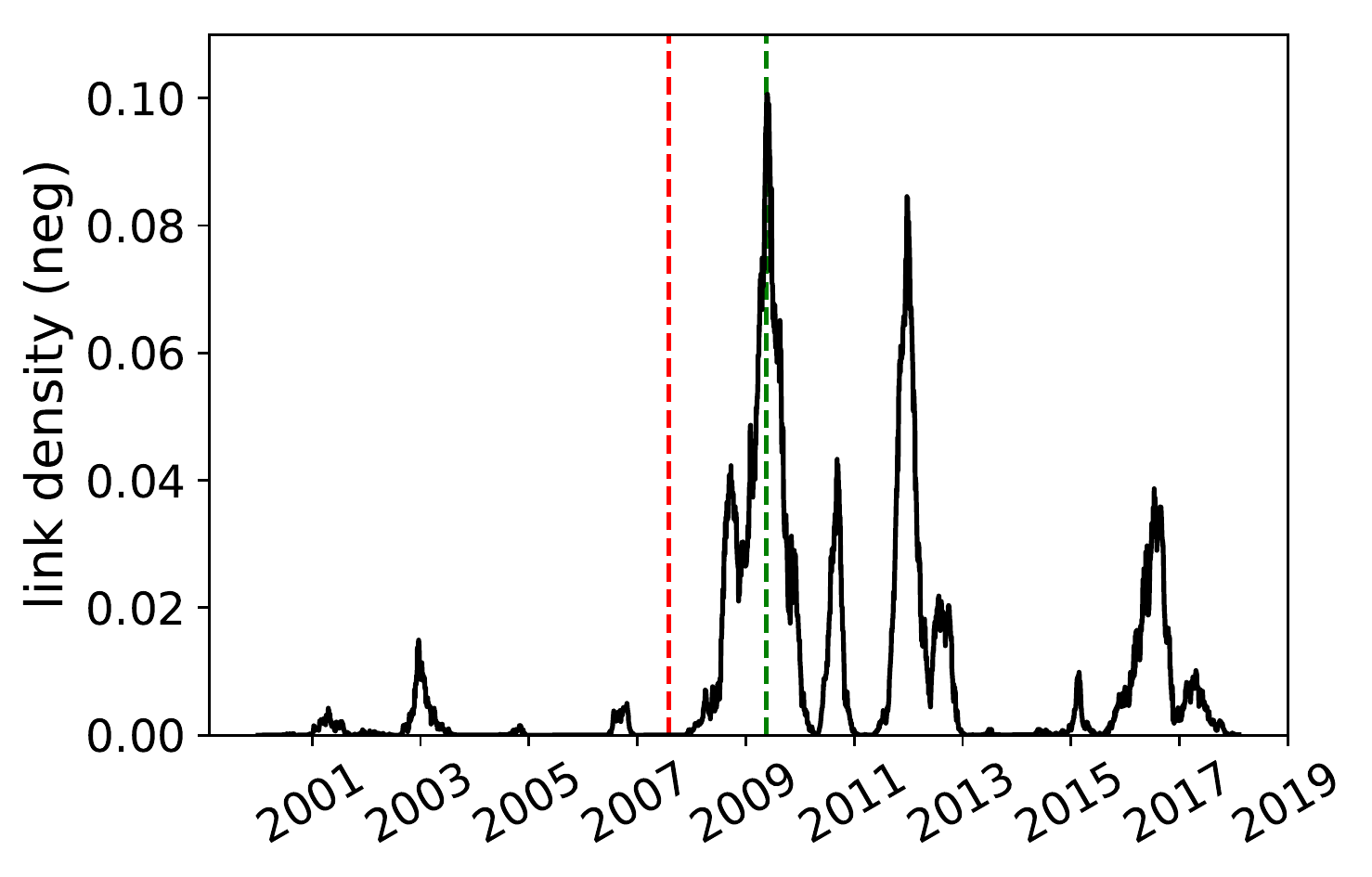}}
\subfigure[\label{fig:AssUE}]{\includegraphics[width=0.4\textwidth]{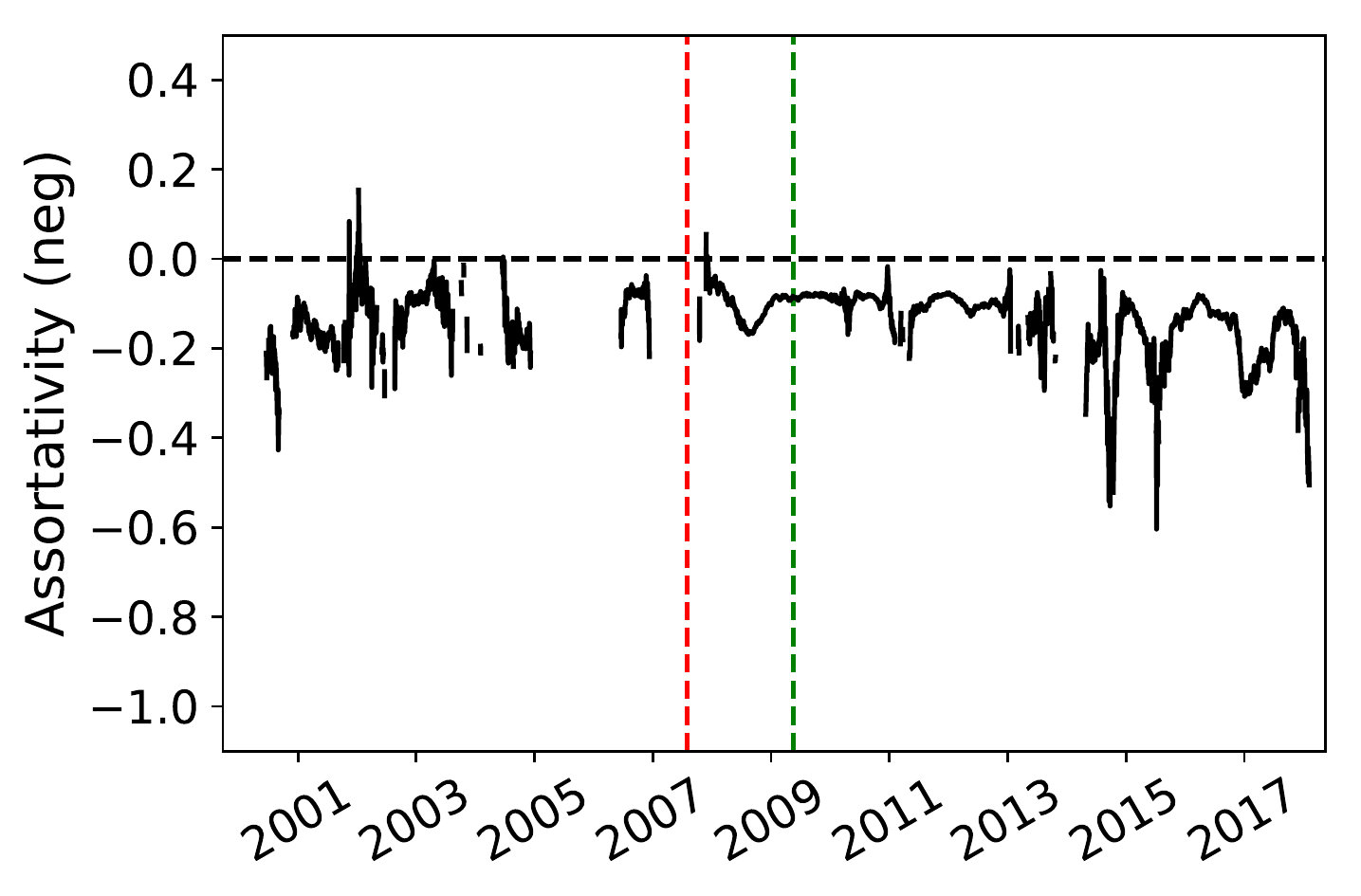}}  

\caption{\small{$(a)$ link density of significant negative $\Phi$; $(b)$ Assortativity of of significant negative $\Phi$ with respect to the sector categorization, only networks with at least $10$ links are considered. }}
\end{figure*}

\end{document}